\documentclass{arxiv}
\usepackage{graphicx}
\usepackage{eurosym}
\usepackage{hyperref}
\usepackage{tablefootnote}

\newcommand{\celsius}{\mbox{$^\circ$C}}
\newcommand{\keuro}{\mbox{k\euro}}
\newcommand{\meuro}{\mbox{M\euro}}
\newcommand{\co}{\mbox{CO$_2$}}
\newcommand{\tco}{\mbox{tCO$_2$}}

\newcommand{\Gtco}{\mbox{GtCO$_2$}}
\newcommand{\coe}{\mbox{CO$_2$e}}

\newcommand{\tcoe}{\mbox{tCO$_2$e}}

\newcommand{\Gtcoe}{\mbox{GtCO$_2$e}}
\newcommand{\tcoeyr}{\mbox{tCO$_2$e yr$^{-1}$}}
\newcommand{\ktcoeyr}{\mbox{ktCO$_2$e yr$^{-1}$}}
\newcommand{\Mtcoeyr}{\mbox{MtCO$_2$e yr$^{-1}$}}
\newcommand{\monetaryef}{\mbox{tCO$_2$e \meuro$^{-1}$}}
\newcommand{\monetaryefyr}{\mbox{tCO$_2$e \meuro$^{-1}$ yr$^{-1}$}}
\newcommand{\massef}{\mbox{tCO$_2$e kg$^{-1}$}}
\newcommand{\massefyr}{\mbox{tCO$_2$e kg$^{-1}$ yr$^{-1}$}}

\newcommand{\aap}{Astron. Astrophys.}   
\newcommand{\asr}{Adv. Space Res.} 
\newcommand{\baas}{Bull. Am. Astron. Soc.}   
\newcommand{\eastro}{Exp. Astron.} 
\newcommand{\nat}{Nature} 
\newcommand{\nastro}{Nat. Astron.} 
\newcommand{\procspie}{Proc. SPIE}   
\newcommand{\pasp}{Publ. Astron. Soc. Pac.}   
\renewcommand\figurename{\bf \small Figure}

\title{Scenarios of future annual carbon footprints of astronomical research infrastructures}

\author{J\"urgen Kn\"odlseder$^{1}$, Mickael Coriat$^{1}$, Philippe Garnier$^{1}$, Annie Hughes$^{1}$}

\begin{document}
\maketitle
\spacing{1.5}

\begin{affiliations}
\item Institut de Recherche en Astrophysique et Plan\'etologie, Universit\'e de Toulouse, CNRS, CNES, UPS,
9 avenue Colonel Roche, 31028 Toulouse, Cedex 4, France
\end{affiliations}

\abstract{
Research infrastructures have been identified as an important source of greenhouse gas 
emissions of astronomical research.
Based on a comprehensive inventory of 1,211 ground-based observatories and space missions, we 
assessed the evolution of the number of astronomical facilities and their carbon footprint from 1945
to 2022.
We found that space missions dominate greenhouse gas emissions in astronomy, showing an
important peak at the end of the 1960ies, followed by a decrease that has turned again into a rise over 
the last decade.
Extrapolating past trends, we predict that greenhouse gas emissions from astronomical
facilities will experience no strong decline in the future, and may even rise substantially, unless 
research practices are changed.
We demonstrate that a continuing growth in the number of operating astronomical facilities 
is not environmentally sustainable.
These findings should motivate the astronomical community to reflect about the necessary evolutions 
that would put astronomical research on a sustainable path.
}
\newline
\normalfont


The unequivocal observation that global warming is caused by human activities leads to the inevitable
question on how these activities need to change in the future.
Faced with a planet that is continuously heating up, resulting in droughts, wildfires, extreme weather 
events, freshwater scarcity, famines, poverty, displacement of populations and geopolitical tensions, 
the consequences of our activities can no longer be ignored.
Scientific research is a common good that contributes to our understanding of the world and the 
anticipation of the future, yet it is increasingly recognised that research activities contribute like all other 
human activities to the deterioration of the climate.\cite{rosen2017}
The ethics committee of the National Centre for Scientific Research (CNRS) in France considers in a
recent opinion\cite{comets2022} that taking into account the environmental impacts of research activities
is an integral part of research ethics, comparable to the respect of humans or experimental animals.
According to the committee, it is the responsibility of scientists to limit the environmental impacts of 
their research practises, but also to consider the impacts of choosing particular research topics and
the means to address them.

Astronomers are particularly active in this introspection, and estimates of greenhouse gas (GHG) 
emissions related to astronomical research activities are becoming an integral part of the scientific 
output of the field.\cite{stevens2020,burtscher2020,jahnke2020,vandertak2021,martin2022,mccann2022,kruithof2023,viole2023}
Primary sources of GHG emissions that were identified include air travelling, purchase of goods and 
services, and use of astronomical facilities such as ground-based telescopes or space observatories. 
It turns out that astronomical facilities are the dominant source of GHG emissions in the field, with an 
estimated average contribution of 36.6$\pm$14.0 \tco\ equivalent (\coe) per astronomer.\cite{knoedlseder2022}
Based on their analysis, ref.~\citen{knoedlseder2022} (hereafter referred to as Paper I) concluded that 
the pace at which new astronomical facilities are built needs to be reduced to make the field sustainable.

In this work we present a quantitative estimate of this conclusion by modelling the future carbon
footprint of astronomical facilities for several scenarios.
We address in particular the question by how much the pace needs to be reduced to limit global
warming to well below 2\celsius\ with respect to pre-industrial levels, as agreed in the Paris Climate 
Agreement.
Based on assessments by the Intergovernmental Panel on Climate Change (IPCC), the United Nations 
Environment Programme, as well as GHG emission reduction targets of the European Space Agency 
and the French Ministry of Higher Education and Research, we translate this ambition into a minimum 
annual carbon footprint reduction of 5\% for astronomical facilities, and a goal of 7\% that is required to 
limit global warming to 1.5\celsius\ (see `Carbon footprint reduction targets for astronomy' in Methods).

We based our work on an inventory of all astronomical facilities that were or are built since 1945 to
model the evolution of their GHG emissions since the end of World War II
(see `Inventory of astronomical facilities' in Methods).
We define a facility as ``astronomical'' if it is (or was) used for astronomical research, 
irrespective of the motivations that led to the realisation of the facility, which may include geopolitical
and economic considerations.
Our inventory comprises 586 ground-based facilities and 625 space missions for which we collected
start and end dates of construction and operations, as well as size, mass and cost information.
Using this information we computed the GHG emission profile of each facility over time, taking
into account emissions from facility construction and operations (see `Computation of annual carbon 
footprint' in Methods).
Similar to Paper I, GHG emissions were computed as a product between activity data and emission factors, 
but in this work both are now time dependent, and reflect the evolution of the activities and the emission 
factors over the years.
In contrast to Paper I, where emissions from construction and operations were integrated over the life cycle 
of the facilities and released uniformly over the life time of facility operations, we used in this work 
a more realistic model that reflects the GHG emissions as they are released over time.
Specifically, construction-related emissions were released during the construction period and 
operations-related emissions during the operating period.
Furthermore, emission factors evolve with the carbon intensity of gross domestic product (GDP) to take 
the gradual decarbonisation of the global economy into account.
As activity data, we used like in Paper I the total cost of construction and the annual cost of operations 
for ground-based observatories, and the total payload mass for space missions.
Aggregating the GHG emission profiles of all facilities, we modelled the history of GHG emissions from 
astronomical facilities over the past 78 years.

We then factorised the GHG emission history into the evolution of the number of active facilities, the 
evolution of the average activity data per facility, and the evolution of the carbon intensity of astronomical 
facilities (see `Future carbon footprint' in Methods).
We fitted the trends of each factor using analytical functions and predicted on their basis extrapolated
trends of future GHG emissions under the hypothesis of a research-as-usual scenario.
We then explored alternative scenarios that either freeze or reduce the number of active facilities,
or that explore scenarios of enhanced decarbonisation, in order to identify pathways that are
compliant with the Paris Climate Agreement.

\section*{Results}

\subsection{Evolution of the carbon footprint of astronomical facilities.}

\begin{figure}[!t]
\centering
\includegraphics[width=16.5cm]{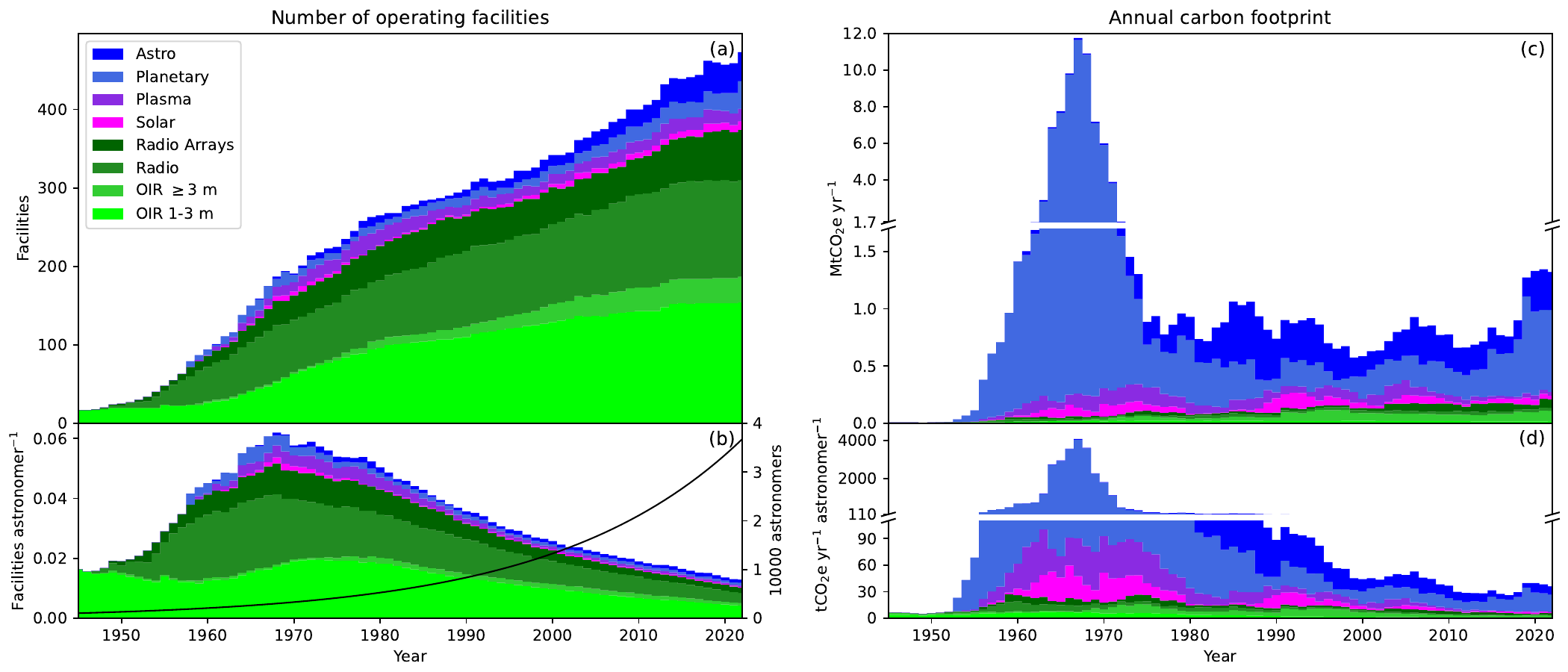}
\caption{
\small
{\bf Evolution of the number of astronomical facilities and their carbon footprint since 1945.}
{\rm
Space missions are represented by blueish colours, distinguishing astrophysics, planetary, plasma physics,
and solar missions.
Ground-based observatories are represented by greenish colours, split into radio arrays, radio telescopes,
and optical-infrared reflectors with diameters between 1-3 m and above.
The number of authors on refereed papers in a given year is used as a proxy for the evolution in the
number of astronomers in the world (solid black line in panel b).
Panel (a) shows the number of operating facilities, panel (b) shows the number of facilities per author, 
panel (c) shows the annual carbon footprint, and panel (d) shows the annual carbon footprint per author.
Note that the y-axes of panels (c) and (d) are split so that the full dynamic range as well as the more
recent evolution is readily visible.
}
\label{fig:evolution}
}
\end{figure}

Figure \ref{fig:evolution} shows the evolution of the number of operating astronomical facilities
and the associated annual carbon footprint since the year 1945.
We qualify as operating all facilities that were in the operations phase in a given year, excluding 
observatories under construction or space missions that were not yet launched, and facilities for 
which operations were terminated.
The number of operating facilities, shown in panel (a), has steadily increased over the years, 
raising from 17 in the year 1945 to 473 in the year 2022.
Ground-based facilities dominate largely the number of operating facilities for all years, 
comprising 85\% and 79\% of the facilities for the years 1980 and 2022, respectively.
For the year 2022, among the 374 operating ground-based facilities, 
41\% were optical-infrared reflectors (OIR) with diameters of 1--3 metres, 
9\% were OIR telescopes with diameters of at least 3 metres, 
33\% were radio telescopes, and 
17\% were radio telescope arrays.
Of the 99 operating space missions in 2022, 
11\% were dedicated to solar physics, 
16\% to plasma physics, 
35\% to planetary exploration, and 
37\% to astrophysics.
Over the last 30 years, the number of operating ground-based facilities has grown by 37\%, 
corresponding to an annual average growth rate of 1.0\% per year, while the number of operating 
space missions has grown by 161\%, corresponding to an annual average growth rate of 3.2\% 
per year.
As the average lifetime of the facilities has not noticeably increased over the last decades
(Supplementary Information), the observed increase in the number of operating facilities is 
attributable to an increase in the deployment pace of new facilities.

Over the same period, the growth rate of the astronomical community was even larger, 
estimated to be 4.6\% per year (Supplementary Information).
This led to a decreasing number of operating facilities that are available per astronomer from 
the early 1970s on (see panel b of Figure \ref{fig:evolution}).
In 2022, there was one facility for every 78 astronomers in operations, while in 1980 a facility
was shared by only 20 astronomers.
Prior to 1970 the trend was opposite, with a tripling of the number of operating facilities
per astronomer between 1950 and 1970 that was mainly driven by the advent of radio astronomy 
and space flights.

The evolution of the annual carbon footprint that we estimated for the facilities in our inventory
is shown in panel (c) of Figure \ref{fig:evolution}.
Note that the y-axis of the panel is broken into two ranges to capture the full dynamics of the 
evolution.
The annual carbon footprint peaks in 1967 at a value of 11.8 \Mtcoeyr\ which is primarily due 
to construction activities related to the missions of the Apollo program.
Excluding these missions, the peak value drops to 3.0 \Mtcoeyr, and excluding all lunar 
exploration missions the peak value drops even further to 1.4 \Mtcoeyr.
One may argue that the motivation of the lunar exploration missions was primarily geopolitical
and not scientific, and hence that these missions should be excluded from the inventory.
Yet, astronomers were profiting from and supporting these missions, hence
their co-responsibility in the associated environmental impacts can not be denied.
Even if lunar exploration missions were excluded, planetary missions still dominate by large the 
annual carbon footprint during the space race, which is attributed to missions towards other solar 
system bodies, including in particular Venus and Mars (Supplementary Information).
This situation lasted until the years 1985--1990 and ended with the collapse of the Soviet 
Union.
From then on the annual carbon footprint is fluctuating between 0.6--1.3 \Mtcoeyr,
dominated to more than 75\% by space missions, and punctuated by the construction of new 
large space facilities, such as the Hubble Space Telescope (HST) or the Compton Gamma-Ray 
Observatory (CGRO) that were launched in the early 1990s.
Over the last decade the annual carbon footprint of space missions is on the rise again, owing 
primarily to planetary exploration missions among which is an increasing number of missions to
the Moon and Mars (Supplementary Information).
In 2022, the annual carbon footprint of astronomical facilities reached a value of 1.3 \Mtcoeyr, of 
which 84\% are attributed to space missions and 16\% to ground-based observatories.
The contribution from space missions originates to 
65\% from planetary exploration missions (corresponding to about half of the total annual emissions in 2022),
30\% from astrophysics missions, and
3\% each from plasma physics missions and solar missions.
The contribution from ground-based observatories originates to 
45\% from OIR telescopes of at least 3 metre diameter,
36\% from radio arrays,
12\% from single radio telescopes, and
8\% from OIR telescopes with diameters between 1--3 metres (totals differ from 100\% due to rounding).
Interestingly, the latter category contributes the largest to the number of operating ground-based 
facilities, but owing to the small size of the observatories, they provide the smallest contribution to 
the carbon footprint.

Dividing the annual carbon footprint by the number of astronomers, panel (d) of 
Figure \ref{fig:evolution} shows that the annual per astronomer footprint decreased from a peak of 
4,079 \tcoeyr\ during the space race to a value of about $\sim$100 \tcoeyr\ in the 1990s.
From there the per astronomer footprint decreased further to a level that remained relatively 
constant over the last decade, comprised within the band 26--40 \tcoeyr, with the lowest and largest 
values reached in 2017 and 2019, respectively.

\subsection{Modelling of the future annual carbon footprints.}

\begin{figure}[!t]
\centering
\includegraphics[width=16.5cm]{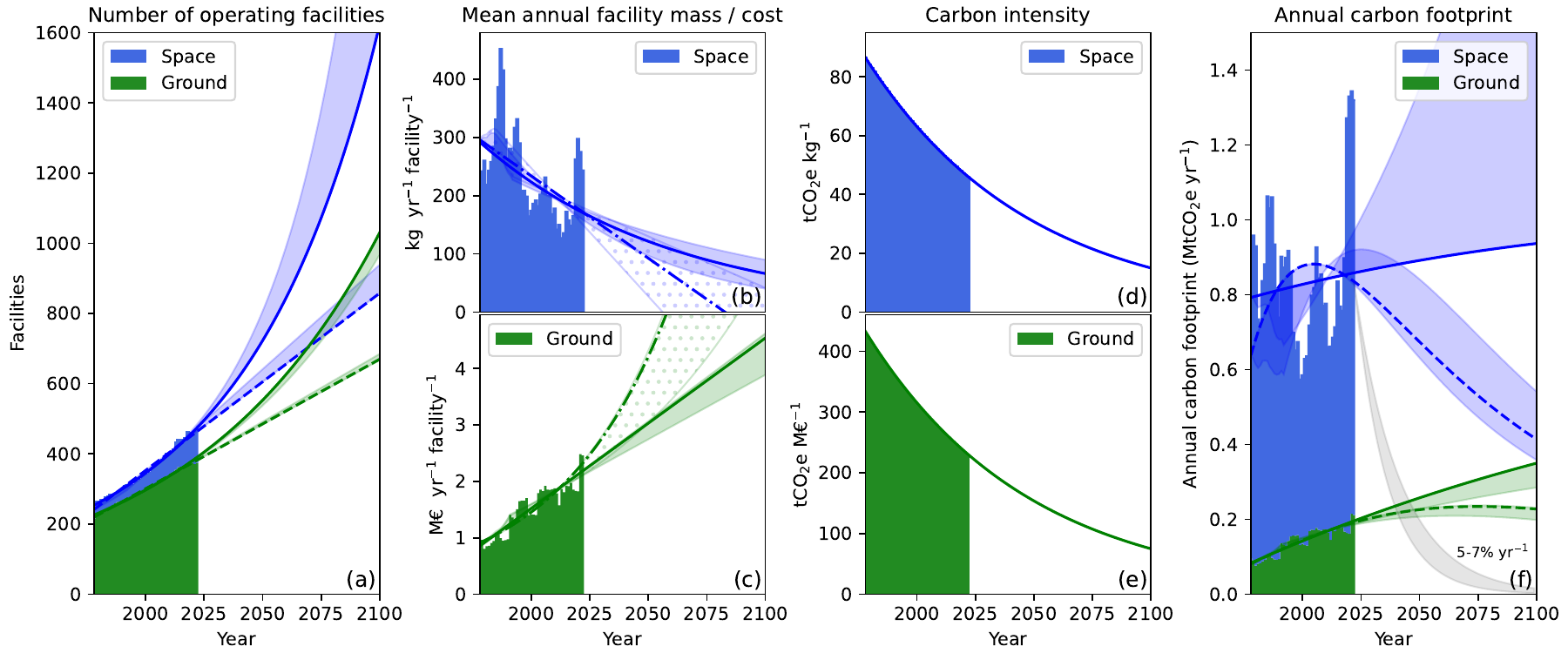}
\caption{
\small
{\bf Modelling of the future annual carbon footprints of astronomical facilities.}
{\rm
Filled histograms represent the data for the past 45 years, with ground-based observatories 
represented in green and space missions in blue.
Panel (a) shows the number of operating facilities,
panels (b) and (c) the mean annual facility mass or cost,
panels (d) and (e) the carbon intensities for space and ground-based facilities,
and panel (f) the annual carbon footprints.
Past trends were adjusted using analytical functions, with shaded bands obtained by
varying the adjustment period from the last 30 to 45 years, and lines corresponding to 
adjustments over the last 45 years.
In panels (a) and (f), solid lines correspond to exponential laws and dashed lines to linear laws 
used for modelling the number of operating facilities.
In these panels, data for space missions are stacked on top of data for ground-based observatories, 
so that the blue histograms, lines and bands represent total numbers and carbon footprints.
In panels (b) and (c), alternative analytical functions that were not used in the analysis are 
shown as dashed-dotted lines and dotted bands for illustration.
In panel (f), the expected trajectory for annual emission reductions of 5--7\% is shown as
grey band.}
\label{fig:projection}
}
\end{figure}

In order to model the future annual carbon footprints of astronomical facilities, we factorised the
latter into the number of operating facilities, the mean annual facility mass or cost, and the carbon 
intensity, with separate decompositions for space missions and ground-based observatories.
For space missions, the mean annual facility mass specifies the total mass of all payloads that 
are constructed in a given year plus 1/48th of all payload masses that are operating in space,
divided by the total number of operating space missions.
The factor of 1/48 reflects the proportion of GHG emissions arising from operations with respect
to construction for a given payload mass (see `Future carbon footprint' in Methods).
For ground-based observatories, the mean annual facility cost corresponds approximately to the 
average annual spendings on ground-based observatories (Supplementary Information).
The carbon intensities specify the GHG emissions as a function of mean annual facility mass or 
cost.
Figure \ref{fig:projection} shows the decomposition of the annual carbon footprint over the last 
45 years, excluding hence the period of the space race.
Past trends were adjusted using either linear or exponential functions to estimate future trends 
until the year 2100, with lines representing adjustments over the last 45 years, and shaded bands 
representing uncertainties that arise from the choice of the adjustment intervals, obtained by 
progressively shortening the interval until it covers only the last 30 years.

Adjusting the growing number of operating facilities using exponential laws over the last 45 years 
suggests annual growth rates of 2.5\% for space missions and 1.2\% for ground-based observatories,
respectively, while linear laws indicate that the number of operating space missions grows by 1.4 per 
year and the number of operating ground-based observatories by 3.7 per year.
We note that data for space missions are better adjusted by an exponential law, while the data for
ground-based observatories favour a linear law.
The mean annual facility mass for space missions decreases somewhat over the years 
while showing important peaks that reflect enhanced construction activities of new
facilities (see Supplementary Information for an interpretation of the figure).
Fitting the data with a linear law predicts implausible negative values before the year 2100,
while an exponential law provides a more plausible description.
Still, the exponential law adjusted over the last 45 years suggests an annual decrease rate of 1.2\% 
with some moderate uncertainty, ending up with a mean annual facility mass in 2100 that is only half 
the value of today.
Although the mean annual facility mass does not directly reflect the payload mass (Supplementary 
Information), an important decrease still implies that satellites will become on average lighter over 
time, a trend that still needs to be confirmed.
In any case, the mean annual facility mass cannot decrease indefinitely, and will plausibly level off 
at some time.
The mean annual facility cost for ground-based observatories rises relatively steadily over the last
45 years and is well explained by a linear law with an increase of 30 \keuro\ yr$^{-1}$ per
facility.
An exponential law fits the data less well, and leads to an import raise of the annual spending per
facility that is probably not compatible with current budget constraints.
Also here, a levelling off in the future seems plausible.
By construction, the carbon intensities are well fit by exponentially decaying laws with decay rates
of 1.44\% (see `Computation of annual carbon footprint' in Methods).

The evolution of the annual carbon footprint, based on the analytical laws that were adjusted to the 
coefficients of the factorisation, is shown in panel (f) of Figure \ref{fig:projection}.
The figure shows as grey band annual emission reductions at a rate of 5--7\% that we assume as
targets for astronomy with the goal to limit global warming to temperatures below 2\celsius\ or
1.5\celsius, respectively (Supplementary Information).
Obviously, the predictions depend on the laws that were used to model the number of operating
facilities, with linear laws leading to lower annual carbon footprints compared to exponential laws.
While the choice of the law impacts the predicted carbon annual carbon footprint for ground-based
observatories only moderately, the situation is different for space missions.
Using a linear law leads to a very curved annual carbon footprint with an emission maximum in the 
years 2004--2012 that actually was not observed.
Despite some strong fluctuations in the historical annual carbon footprint data, the data are better 
explained when an exponential law is used, reproducing the gradually increasing trend in the annual
carbon footprint.
In any case, for both laws the predicted annual carbon footprint is well above the suggested 
emission reduction trajectory for astronomy.

\subsection{Future annual carbon footprints.}

\begin{figure}[!t]
\centering
\includegraphics[width=16.5cm]{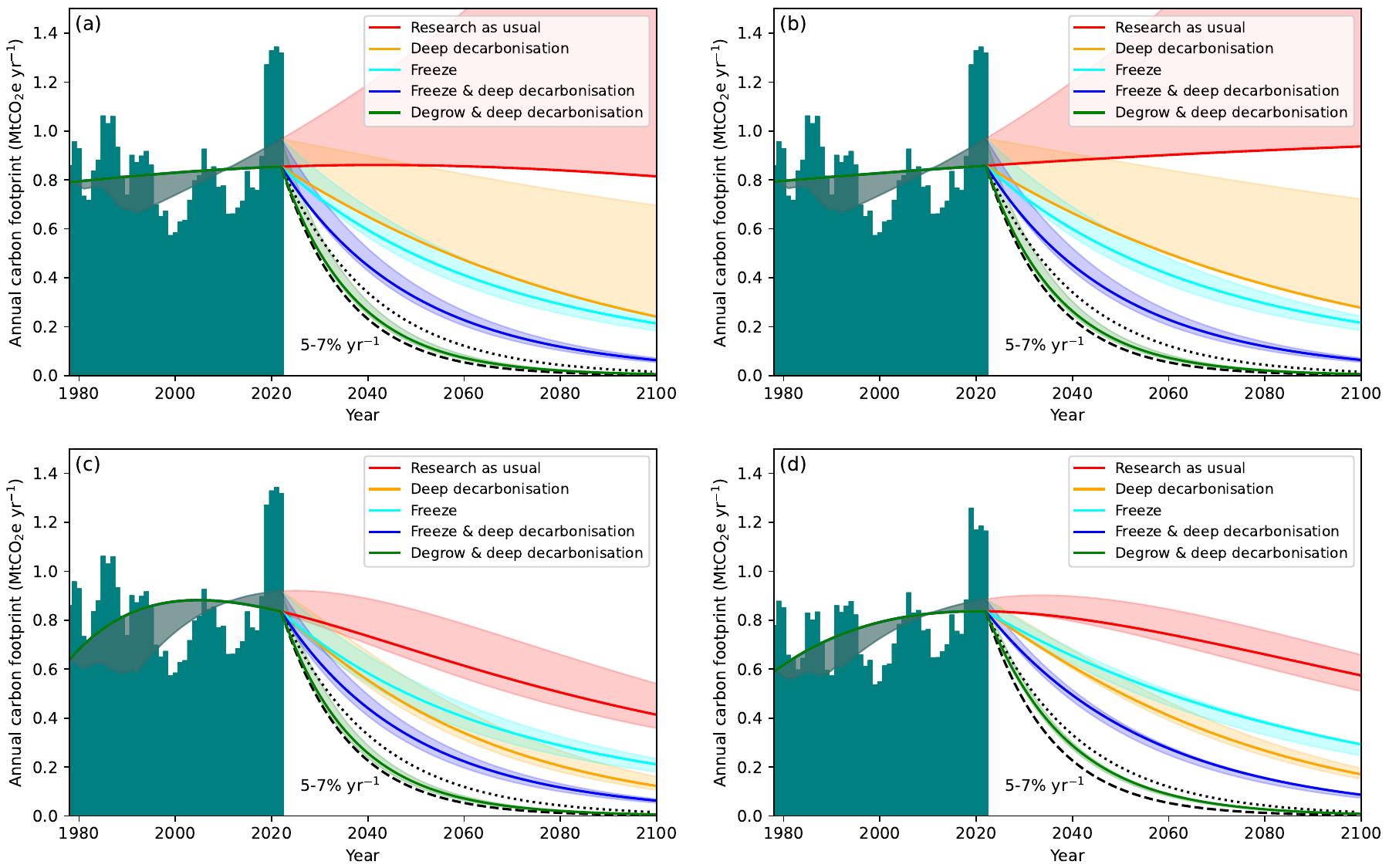}
\caption{
\small
{\bf Future annual carbon footprints for five illustrative scenarios.}
{\rm
The `Research as usual' scenario corresponds to the projections from the carbon footprint factorisation.
In the `Deep decarbonisation' scenario, the decarbonisation rate after the year 2022 was increased 
from 1.44\% to 3\% per year.
In the `Freeze' scenario the number of operating facilities was kept constant at the value of the year 
2022, and both measures were combined in the `Freeze \& deep decarbonisation' scenario.
In the `Degrow \& deep decarbonisation' scenario, the number of operating facilities was
decreased from value in 2022 at the rate of 3\% per year.
Each panel corresponds to a different modelling of the factorisation coefficients.
For panel (a), the number of operating facilities was modelled using an exponential law for
space missions and a linear law for ground-based observatories.
In panel (b), exponential laws were used in both cases, while in panel (c) linear laws were employed.
For panel (d), in addition to using linear laws, payloads with masses inferior to 20 kg and superior 
to 10 tons were excluded in the adjustments.
Shaded bands were obtained by varying the adjustment period of the analytical laws from the last 30 
to 45 years, lines corresponding to adjustments over the last 45 years.}
}
\label{fig:scenarios}
\end{figure}

In order to investigate possible lever arms for emission reductions we explored five illustrative 
scenarios (see `Future carbon footprint' in Methods).
Figure \ref{fig:scenarios} shows the predicted annual carbon footprint of these scenarios for 
different modellings of the factorisation coefficients, where panel (a) represents the modelling
that reproduces most accurately the historic trends (Supplementary Information).
In the `Research as usual' scenario, where no additional measures with respect to current practices
are taken, the extrapolation of the historic trends over the last 45 years suggests a relatively constant
annual carbon footprint until the end of the century.
According to our modelling, the growth in the number of facilities is in this case in balance with the 
decarbonisation of the industry and the suggested reduction of the mean annual payload mass, 
stabilising the annual carbon footprint in the long run.
Estimating the historic trends over shorter periods suggests however that the annual carbon 
footprint may also increase over time, reaching a value that is at least twice as large in 2100 
compared to current levels.
In the `Deep decarbonisation' scenario, where we increased the decarbonisation rate from 1.44\% 
to 3\%, corresponding to a doubling of current efforts, the annual carbon footprint in 2050 is reduced 
by 11--35\% with respect to 2022 levels.
Alternatively, preventing a rise in the number of operating facilities by keeping the fleet of 
facilities at the level 2022, as explored in the `Freeze' scenario, bears a reduction potential 
of 41--47\% with respect to 2022 levels.
Combining the stabilisation of the number of operating facilities with an enhanced decarbonisation,
explored by the `Freeze \& deep decarbonisation' scenario, promises emission reductions of
62--66\% with respect to 2022 levels.
Still, none of the scenarios meets the annual carbon footprint reduction target, which is situated
in 2050 at 76--87\% with respect to 2022 levels for the targeted annual decrease rates of 
5--7\% per year.
We therefore went one step further and decreased the number of operating facilities in the
`Degrow \& decarbonisation' scenario by a constant rate of 3\% per year, combined with
a decarbonisation rate of 3\% per year, which finally is a scenario that is compliant with a trajectory
that keeps global warming to below 2\celsius.
We repeated the exercise for alternative modellings of the factorisation coefficients (panels c--d)
with always reaching the same conclusion: the only scenario that fits a trajectory for keeping
global warming to below 2\celsius\ is the `Degrow \& decarbonisation' scenario 
(Supplementary Information).

\section*{Discussion}

By estimating the carbon footprint of astronomical facilities for the year 2019, Paper I revealed that they 
were the single largest contributor to the professional GHG emissions of an astronomer.
While Paper I was based on a list of 85 facilities from which the global impact of the world-fleet was 
extrapolated, we assessed in this work the global carbon footprint without any extrapolation.
For the year 2019, for which Paper I estimated an annual carbon footprint of 36.6$\pm$14.0 \tcoeyr\ per 
astronomer, our analysis suggests an average annual footprint of 39.7 \tcoeyr\ per astronomer.
Although this is in good agreement with Paper I, we attribute 87\% of the emissions in 2019 to space 
missions, while Paper I concluded that 62\% of the emissions originate from ground-based observatories.
We explain these differences by an overestimation of GHG emissions from ground-based observatories 
in Paper I due to a bias towards large telescopes in the extrapolation procedure, combined with an 
underestimation of GHG emissions from space missions due to the uniform distribution of life cycle 
emissions over the operating life time (Supplementary Information).
As can be seen in panel (b) of Figure \ref{fig:evolution}, the annual carbon footprint fluctuates substantially
over time, owing to the construction activities of new facilities, which introduces a large variability in the
annual carbon footprint that was smoothed out by the method used in Paper I.

This underlines the importance of studying the full emission history, so that conclusions can be drawn from
long-term trends, and not from particular fluctuations that may occur in a specific year.
Studying the long term trends provides also insights into future GHG emissions, and our analysis 
suggests that they probably will not decline in a research-as-usual scenario, despite the assumption that 
the current decarbonisation trends will persist.
The reason behind the maintenance of GHG emissions at current levels, if not their rise above them, is that
the current growth rate of astronomical facilities largely compensates present decarbonisation efforts.
The main open question is whether this growth will be linear or exponential, with eventually a modest 
gradual decline of emissions in case of a linear growth.
Yet, even if emissions decline, the rate will be much too slow, and GHG emissions will be well above the 
required trajectory for limiting global warming to well below 2\celsius.

So how can GHG emissions from astronomical facilities be reduced?
One option is enhancing decarbonisation, by massively investing in renewable energy sources, using 
bio-sourced and recycled materials, reducing any use of fossil fuels, and limiting the dimensions and 
weights of future facilities.
Probably this will reach limits, in particular for the space sector, since rocket launches are hard to abate.
We explored a scenario where current decarbonisation trends are doubled, but this fails largely to meet
the required trajectory.
Annual decarbonisation of the economy would need to be increased from 1.4\% to 7--9\% per year
for the next decades to allow the continued growth of astronomical facilities while meeting the required 
trajectory.
These levels are only reached in the most ambitious C1 scenario of the IPCC\cite{ipcc2022wg3} that relies 
heavily on carbon capture and storage (CCS) and carbon dioxide removal (CDR) technologies that should 
remove on average 10 \Gtco\ per year (about 25\% of the current annual \co\ 
emissions\cite{friedlingstein2022}) over the next 80 years, which seems to be highly 
unrealistic.\cite{iea2022,ho2023,editorial2024,dziejarski2023}
A safer more rational solution would be to stop the growth in the number of operating facilities by keeping
it at current levels.
This would not mean that no new astronomical facilities would be developed, but only existing facilities
would be replaced at the end of their life time by facilities of equivalent mass or cost.
If more heavy or expensive facilities would be desired (in the sense of the activity data used in this work),
the end of life of several outgoing facilities would need to be waited out before new facilities could be put
in operations.
Freezing the number of astronomical facilities to current levels would reduce the decarbonisation needs to 
5--7\% per year in order to meet the required trajectory.
This level is comparable to, but slightly larger than, the emission reductions that were reported due to 
COVID-19 lockdowns in 2020\cite{jackson2022}, which hence seems to be still very ambitious, but eventually
feasible.
However, if this level turns out to be unfeasible, the only remaining solution is to decrease the number of 
operating facilities.
In Figure 3 we used the example of an annual decrease of 3\% per year, which would relax the 
decarbonisation requirements to 2--4\% per year, a level that is probably achievable with some efforts.

The situation is hence clear: there is no evidence for the possibility of a `green growth' of astronomical 
facilities, which is actually no surprise, as others have evidenced the same for the economy 
at large.\cite{jackson2019,hickel2020,hannesson2021,murphy2022,vogel2023}
We therefore call the astronomical community to reflect on this fact, and to engage into a debate on how
astronomical research can be made sustainable.
A way forward would be to include sustainability requirements in competitive grant processes\cite{mass2022}
and to assess environmental impacts of new research facilities prior to taking implementation 
decisions, pondering carefully over the trade-offs between environmental impacts, community interest, and 
scientific return.
Basing any new facility on a collaboration of the world-wide astronomical community, and hence abandoning 
the logic of international competition, appears also fundamental, as the latter often leads to the unnecessary 
multiplication of similar facilities.
Furthermore, large volumes of archival data from past and existing facilities remain poorly explored, hence
the maintenance and exploration of these treasures should become a priority, and attract more attention and 
resources.
Requesting that each research domain fully exploits any existing archival data before engaging into the 
implementation of a new facility does not seem to be outrageous.
The community may also question whether astronomical research should continue to serve as a pretext for 
satisfying geopolitical interests and economic expansion, with the rising interest for lunar exploration and
exploitation being a typical example.\cite{flahaut2023,schneider2024}

Astronomy is of course not a singularity, and any scientific field that relies on large facilities will probably 
come to similar conclusions, once the communities engage in a comprehensive evaluation of their 
environmental impacts.
Astronomy, as many other research fields, creates unquestionably many benefits to society, and it is
finally a societal and political choice to decide how much environmental impacts are tolerated for these
gains. 
Astronomy distinguishes itself by a unique perspective, as it adresses questions about our origins and our 
place in this world, questions that bear profound cultural implications.
Astronomers know best that there is no `Planet B', and that the only realistic option for humanity is to 
preserve liveable conditions on Earth.
We believe that astronomers have therefore also the ethical responsibility to lead by example, showing 
how these conditions can be maintained for the generations to come, without compromising human 
curiosity and excitement about the world in which we are living.

\vspace{12pt}
\begin{methods}

\subsection{Carbon footprint reduction targets for astronomy.}

Although there exist currently no agreed carbon footprint reduction targets for astronomy, the 
IPCC reminds in the sixth assessment report that limiting warming to 1.5\celsius\ or 2\celsius\ ``involve[s] 
rapid and deep and in most cases immediate greenhouse gas emission reductions in all 
sectors''.\cite{ipcc2022wg3}
As has been shown\cite{knoedlseder2024}, astronomy is a sector that contributes to global warming, 
and consequently it is subject to this imperative.
The IPCC reminds that ``doing less in one sector needs to be compensated by further reductions in 
other sectors if warming is to be limited''.\cite{ipcc2022wg3}
Eventually, policy makers and the society at large may set different reduction efforts for different activity
sectors, yet until this is done, a reasonable assumption is that astronomy needs to reduce GHG emissions
at the same pace as the humanity at large.

According to the 2023 Emission Gap Report of the United Nations Environment Programme (UNEP),
global GHG emissions increased by 1.2\% from 2021 to 2022 to reach a new record of 57.4 
gigatons of \co\ equivalent (\Gtcoe).
Least-cost pathways that are consistent with limiting global warming to 2\celsius\ or 1.5\celsius\ imply
that GHG emissions need to be reduced to 20 \Gtcoe\ or 8 \Gtcoe\ in 2050, respectively, corresponding
to annual reductions of 3.7\% or 6.8\% per year.
It should be recognised that these are average global reduction targets, and that the targets should be 
higher for developed countries owing to their larger responsibility in global warming, a principle that was
agreed upon in the Paris Climate Agreement.

There do actually exist some carbon footprint reduction targets in fields related to astronomy, which may
also server as guideline when discussing targets for astronomy.
For example, at the ministerial Council meeting of the European Space Agency (ESA) on 22 November 
2022, it was decided to reduce ``the carbon footprint of the Agency by 46\% by 2030 as compared to 
the 2019 baseline''\cite{esa2022}, which corresponds to annual reductions of 5.4\% per year over the 
considered period.
The Ministry of Higher Education and Research in France defines an annual carbon footprint reduction 
goal of 5\% for all French research organisations that applies also to astronomical research in that 
country.\cite{mesr2022}

We therefore required as the strict minimum an annual carbon footprint reduction of 5\% for astronomical
facilities in our study, with the goal of reaching an annual reduction of 7\% that is required 
to limit global warming to 1.5\celsius.
We note that the resulting 1.5\celsius\ pathway is very similar to the cross sector net-zero pathway
of the Science-Based Targets Initiative\cite{sbti2021}.

\subsection{Inventory of astronomical facilities.}

Our study is based on an inventory of astronomical ground-based observatories and space missions 
that were constructed or active in the period from 1945 to 2022.
We initiated our inventory of ground-based observatories from the list of active professional
observatories compiled by Lukac \& Miller \cite{lukac2000}, complemented by information
gathered from literature, including in particular topical 
books\cite{leverington2017,thompson2017,baars2018,orchiston2021,kellermann2021},
reviews, observatory reports, public websites, and press articles.
For each facility we compiled, among other data, information on years of operation start
and end, and, if available, start of construction, as well as construction and annual operation
costs.
Cost data were inflation corrected to 2019 economic conditions in their quoted currencies
based on national consumer price indices (CPI), and converted to 2019 Euros using purchasing
power parities (PPP) provided by the Organisation for Economic Co-operation and Development 
(OECD).
Unfortunately, cost data are only sparsely publicly available, hence we use cost-diameter
scaling relations to infer construction and annual operation costs from the facilities' diameters
in case of lacking cost information (Supplementary Information).
For that purpose we collected the facilities' reflector or antenna diameters, or in case of 
non-circular geometries, converted the facilities' collection areas $A$ into effective diameters
$D$ using $D=2\sqrt{A/\pi}$.
Public information on construction start dates is also sparse, and in case of lacking information
we assumed that construction started five years before operations.
Furthermore, we identified in each category the type of the facility so that we could for
example distinguish optical-infrared reflectors (OIR) with monolithic thick mirrors from those 
with segmented mirrors, and apply dedicated scaling relations based on the technology
employed.
Finally, we classified ground-based observatories into four categories: 
large-sized (diameter $\ge3$ metre) and
medium-sized (diameter $1-3$ metre) OIRs,
radio telescopes, and radio telescope arrays.
Our inventory comprises 586 past, active or planned ground-based facilities,
including
43 large-sized OIR telescopes,
202 medium-sized OIR telescopes,
215 radio telescopes, and
126 radio telescope arrays.
Although we aimed for completeness from the year 1945 onwards, we cannot exclude that a 
few facilities, in particular small radio telescopes or medium-sized OIR telescopes, have been
missed.

We based our inventory of space missions on the comprehensive list provided on Gunter's space 
page\cite{krebs2023}, complemented by information gathered from public websites, including those 
of space agencies, projects, and research institutes, and press articles.
For each mission we compiled, among other data, information on years of mission launch and 
mission end, if available also start of mission phases A/B (feasibility and preliminary definition)
and phases C/D (detailed definition, qualification and production), as well as payload launch (wet)
mass, and, if available, construction, operations and full mission costs.
We classified space missions according to their scientific purpose, separating missions for
solar observations, studies of interplanetary plasma physics, planetary exploration, and
astrophysics.
Our inventory comprises 625 past, active or planned space missions since the beginning of the
space age (1957), including 
61 solar missions,
130 plasma physics missions,
279 planetary missions, and
155 astrophysics missions.

\subsection{Computation of annual carbon footprint.}

Similar to Paper I, we define the carbon footprint of an astronomical facility as the aggregate of all 
greenhouse gas (GHG) emissions generated by both the construction and the operations of the facility.
To enable aggregation of the different gases that cause global warming, GHG emissions are expressed 
in carbon dioxide equivalents (\coe) which take into account the different warming potentials and lifetimes 
of the various GHGs in the atmosphere. 
Carbon footprints were computed by multiplying activity data with emission factors.
For ground-based observatories we used construction and annual operation costs as activity data, while 
for space missions we based our estimates on the payload launch mass.
Paper I derived emission factors for astronomical facilities from the literature, finding 
240 \monetaryef\ for the construction and 250 \monetaryefyr\ for the operations of ground-based 
observatories.
For space missions, Paper I discussed two options, one using total mission cost and another using 
payload launch mass as activity data, showing that when GHG emissions are aggregated, both proxies 
lead to equivalent results.
Since payload launch masses are readily available for all space missions and not subject to significant
uncertainties, we used them as activity data in our study (an alternative analysis using
mission cost as activity data is available in the Supplementary Information).
While Paper I could only derive a life-cycle emission factor of 50 \massef\ on the basis of two cases 
studies\cite{wilson2019}, more detailed information for the two cases studies were provided recently 
in ref.~\citen{wilson2023}, suggesting emission factors of 48 \massef\ for construction and 1 \massefyr\ 
for operations.
We adopt these factors for our study, so that the results can be directly compared to those of Paper I
(Supplementary Information).

Since the publication of Paper I, more estimates for emission factors for astronomical facilities 
have become available, and they confirm that the adopted emission factors are of the right order of magnitude.
Ref.~\citen{kruithof2023} determined the carbon footprint of the LOFAR telescope and found
emissions factors of 100 \monetaryef\ for construction and 187 \monetaryefyr\ for operations, 
emphasising that these were lower limits since some processes were excluded in their
analysis.
Using a comprehensive Life Cycle Assessment (LCA) of one Mid-Sized Telescope of the Cherenkov 
Telescope Array, ref.~\citen{dossantosilla2024} found emission factors of 551 \monetaryef\ for the 
construction of the telescope structure and 70 \monetaryef\ for the construction of the camera, 
resulting in 311 \monetaryef\ for one telescope.
This suggests that construction of different subsystems have different emission factors, but when 
combined the resulting global emission factor comes reasonably close to the one adopted in our work.
Finally, applying a LCA to the X-IFU instrument on ESA's Athena space mission, ref.~\citen{barret2024}
determined an emission factor of 115 \tcoe\ per kg of instrument mass for instrument development,
which is considerably driven by the subtantial labour that is needed to develop the instrument.

The emission factors are expected to decrease over the years due to reductions in the carbon 
intensity of energy production (for example as result of replacing fossil fuels by renewable 
energies) and due to improvements in the energy efficiency of construction processes and
operation activities.\cite{mccann2022}
Since quantitative estimates of these evolutions do not exist for astronomical facilities, 
we assumed that the emission factors follow the general decarbonisation trends
that are observed for the economy at large.
The latter are readily traced by the carbon intensity of inflation corrected GDP based on PPP, 
which we used as a proxy for the time dependence of the emission factors.
As shown in the Supplementary Information, the observed trends are well described over the
period of interest by the piecewise function
\begin{equation}
CI(t) = 2.5 \times \left\{
\begin{array}{@{}ll@{}}
0.72 - 0.00113 \times (t-2019)& \mbox{for}\,\,t\le1973 \\
0.40 \times \exp\left(-0.0144 \times(t-2019)\right) & \mbox{else}
\end{array}
\right.
\label{eq:carbon-intensity}
\end{equation}
where $t$ is time given in years.
The function applies to economic conditions of 2019 and is normalised to $CI(2019)=1$ so
that it can be directly multiplied with the adopted emission factors that were derived for the
same reference year.

To determine the annual carbon footprint $F(t)$ of each facility we assumed that construction 
emissions occur over a period $t_c \le t < t_o$, where $t_c$ is the year of construction start 
and $t_o$ is the year of operations start or mission launch.
Furthermore, we assumed that emissions from operations occur subsequently over a period 
$t_o \le t \le t_e$, where $t_e$ is the year when operations end, or 2022 as the end year of
the considered period, whichever is smaller.
Finally, costs and payload masses used as activity data for construction were annualised 
by dividing them by $t_o -t_c$, so that they were equally distributed over the construction
period.
We then computed the annual carbon footprint $F(t)$ in units of \tcoeyr\ for each ground-based 
observatory using
\begin{equation}
F(t) = CI(t) \times \left\{
\begin{array}{@{}ll@{}}
240 \times \frac{\mbox{Construction costs (\meuro)}}{\mbox{$t_o -t_c$}} & \mbox{for}\,\,t_c \le t < t_o \\
250 \times \mbox{Annual operating costs (\meuro)} & \mbox{for}\,\,t_o \le t \le t_e \\
0 & \mbox{else}
\end{array}
\right.
\label{eq:ground-footprint}
\end{equation}
and for space missions using
\begin{equation}
F(t) = CI(t) \left\{
\begin{array}{@{}ll@{}}
48 \times \frac{\mbox{Payload mass (kg)}}{\mbox{$t_o -t_c$}} & \mbox{for}\,\,t_c \le t < t_o \\
1 \times \mbox{Payload mass (kg)} & \mbox{for}\,\,t_o \le t \le t_e \\
0 & \mbox{else}
\end{array}
\right.
\label{eq:space-footprint}
\end{equation}
over the period $1945 \le t \le 2022$.
Illustrations of $F(t)$ for one ground-based observatory and one space mission are given in the 
Supplementary Information.

\subsection{Evolution of the number of astronomers.}

To compare the evolution of the world-fleet of astronomical facilities and the associated
carbon footprint to the size of the astronomical community, we determined the latter using the publication 
database of the Astrophysics Data System (Supplementary Information).
We used the number of individual authors affiliated to astronomical research institutes that signed 
publications in refereed journals as a proxy of the size of the astronomical community, and extracted 
its evolution for the period 1940--2022.
We verified using literature data that our estimates match independent assessments of the size of 
the astronomical community.\cite{knoedlseder2022,jascheck1991}
From the year 1975 on, where the extracted data do not suffer from incompleteness, the number of
astronomers evolves exponentially, with a doubling time of 15 years.
The data closely follow the relation
\begin{equation}
\mbox{\it Number of individual authors}(t) = 32018 \times \exp\left(0.046 \times(t-2019)\right)
\label{eq:authors}
\end{equation}
that we adopt for the evolution of the number of astronomers in our study.

\subsection{Future carbon footprint.}

To estimate the future carbon footprint of astronomical facilities, we factorised the annual footprints of 
ground-based observatories and space missions using a formula that was inspired by the Kaya identity, 
which is frequently used in the development of future GHG emission scenarios.\cite{kaya1997}
The Kaya identity is a mathematical identity that expresses GHG emissions as the product of human 
population, GDP per capita, energy intensity (per unit of GDP), and carbon intensity (emissions per unit 
of energy consumed).
We use a similar approach to model the evolution of the carbon footprint of astronomical facilities 
by decomposing the footprint into a number of facilities, mean mass (for space missions) or cost (for 
ground-based observatories) per facility, and carbon intensity per unit of mass or cost.
Specifically we used
\begin{equation}
F(t) = N(t) \times \frac{C(t)}{N(t)} \times \frac{F(t)}{C(t)}
\label{eq:kaya}
\end{equation}
where
$N(t)$ is the number of facilities, and $C(t)$ is the activity data that when multiplied with the carbon
intensity $F(t)/C(t)$ provides the annual carbon footprint of the facilities.
For ground-based observatories, $C_g(t)$ was computed by summing over individual observatories $i$
using
\begin{equation}
C_g(t) = \sum_i \left\{
\begin{array}{@{}ll@{}}
\frac{\mbox{Construction costs of observatory $i$ (\meuro)}}{\mbox{$t_o^i -t_c^i$}} & \mbox{for}\,\,t_c^i  \le t < t_o^i  \\
\frac{250}{240} \times \mbox{Annual operating costs of observatory $i$ (\meuro)} & \mbox{for}\,\,t_o^i  \le t \le t_e^i  \\
0 & \mbox{else}
\end{array}
\right.
\label{eq:ground-spending}
\end{equation}
and corresponds to the annual spending for observatory construction plus 250/240 times the annual 
spending for observatory operations.
The factor 250/240 takes into account the different emission factors for construction and operations of 
ground-based facilities, putting both contributions on the same scale so that a single carbon intensity 
$F_g(t)/C_g(t)$ could be used that corresponds to the emission factor for observatory construction.
$C_g(t)$ corresponds roughly to the annual spending on ground-based observatories (Supplementary 
Information).
In analogy, $C_s(t)$ was computed for space missions using
\begin{equation}
C_s(t) = \sum_i \left\{
\begin{array}{@{}ll@{}}
\frac{\mbox{Payload mass of mission $i$ (kg)}}{\mbox{$t_o^i -t_c^i$}} & \mbox{for}\,\,t_c^i \le t < t_o^i \\
\frac{1}{48} \times \mbox{Payload mass of mission $i$ (kg)} & \mbox{for}\,\,t_o^i \le t \le t_e^i \\
0 & \mbox{else}
\end{array}
\right.
\label{eq:space-payload}
\end{equation}
where the factor 1/48 is the ratio between the emission factors for operations and construction,
so that the carbon intensity $F_s(t)/C_s(t)$ for space missions corresponds to the emission factor
for satellite construction.
$C_s(t)$ corresponds thus to the sum of the payload masses that are constructed in a given
year plus 1/48th of all payload masses that are operating in space.

We then modelled each factor of Equation \ref{eq:kaya} by adjusting either linear or exponential 
functions to the past evolutions.   
To avoid the singularly large carbon footprints associated to the initial space race, we limited the
adjustments to the years from 1978 onwards.
We performed an initial adjustment of the functions over the years 1978--2022, and then 
successively increased the start year up to 1993 to assess the uncertainty related to the particular 
choice of an adjustment period.
In that way, the functional parameters of each term of Equation \ref{eq:kaya} was estimated from 
the observed trends over the past 30--45 years.

We then explored five possible scenarios for the evolution of the future carbon footprint of astronomical
facilities.
First we explored a `Research as usual' scenario, where $N(t)$ and $C(t)/N(t)$ evolve according
to the adjusted functional laws, and $F(t)/C(t)$ decreases exponentially with a fixed rate of 1.44\% per 
year, corresponding to the historical decrease of the carbon intensity (see Eq.~\ref{eq:carbon-intensity}).
Second, we explored a `Deep decarbonation' scenario where the decarbonisation rate was increased 
from 1.44\% to 3\% per year, doubling the current efforts in the reducing the carbon footprint of
the economy.
Third, we explored a scenario dubbed `Freeze', where we fixed the number of operating facilities to 
the one in the year 2022.
This corresponds to a scenario where the number of operating facilities would be frozen at the 
current level, implying that only observatories or missions that end operations would be replaced by 
new facilities.
We combined the freezing and the deep decarbonisation in a fourth scenario called 
`Freeze \& deep decarbonisation' to explore the additional effects of both measures.
Finally, we explored a scenario dubbed `Degrow \& deep decarbonisation', where in addition
to the decarbonisation rate of 3\% per year the number of operating facilities is decreased by
3\% per year with respect to the number in the year 2022.
It turned out that this level of decrease is needed to meet the proposed GHG emission reduction
targets for astronomy.

\end{methods}

\section*{Data Availability}

All data used for this work are available for download at \url{https://zenodo.org/records/12568160}\cite{zenodo}.

\section*{Code Availability}

All code used for this work is available for download at \url{https://zenodo.org/records/12568160}\cite{zenodo}.

\begin{addendum}

\item[Correspondence] 
Correspondence should be addressed to J.K. (jurgen.knodlseder@irap.omp.eu).

\item
This research has made use of NASA's Astrophysics Data System Bibliographic Services (ADS).
We thank K.~Lockhart for her help with using that service.
We further would like to than
D. Barret
for useful discussions.
This work has also benefited from discussions within the research group and cross-disciplinary 
collective Labos~1point5, and within the grassroots movement Astronomers for Planet Earth.
We thank all members for their engagement and support.
This work has made use of the Python 2D plotting library matplotlib\cite{hunter2007}.

\item[Author Contributions Statement]

J.K. developed the method, gathered and analysed the data, and drafted the paper.
P.G. helped with the collection of some data.
All authors contributed to the definition of the analysis method, the elaboration of the discussion section,
and the review of the manuscript.

\item[Competing Interests]
The authors declare no competing interests.

\item[Peer review information]

\item[Acronyms]
Acronyms used throughout this paper represent:
Carbon Capture and Storage (CCS),
Carbon Dioxide Removal (CDR),
Compton Gamma-Ray Observatory (CGRO),
Consumer Price Index (CPI),
European Space Agency (ESA),
Greenhouse Gas (GHG),
Gross Domestic Product (GDP),
Hubble Space Telescope (HST),
Intergovernmental Panel on Climate Change (IPCC),
Life Cycle Assessment (LCA),
Low Frequency Array (LOFAR),
National Centre for Scientific Research (CNRS),
Optical Infrared Reflector (OIR),
Organisation for Economic Co-operation and Development (OECD),
Purchase Power Parity (PPP),
United National Environment Programme (UNEP),
X-ray Integral Field Unit (X-IFU).

\end{addendum}


\newpage
\renewcommand\tablename{\bf \small Supplementary Table}
\renewcommand\figurename{\bf \small Supplementary Figure}
\setcounter{table}{0}
\setcounter{figure}{0}
\setcounter{equation}{0}
\section*{\Large Supplementary Information}

\section*{Cost scaling relations for ground-based facilities}

To cope with lack of publicly available cost data for ground-based facilities we made use of scaling 
laws between telescope collection diameter $D$ and cost $C$.
Such scaling laws have been discussed in the literature, and in their simplest form assume that the cost 
varies in proportion to some power $n$ of the diameter
\begin{equation}
\mbox{C} = k \times \left( \frac{D}{D_0} \right)^{n}
\label{eq:cost}
\end{equation}
with $n$ comprised between 1.7 and 2.8.\cite{supp:hoerner1975,supp:meinel1982,supp:humphries1984,supp:stepp2003,supp:vanbelle2004,supp:stahl2019}
We fitted Supplementary Equation \ref{eq:cost} to our construction cost and diameter data to derive 
scaling laws for different telescope categories and types.
The resulting parameters are summarised in Supplementary Table \ref{tab:scalingrelations},
with quoted uncertainties reflecting the mean absolute deviation of construction costs from the
scaling laws.
The diameter and construction cost data are shown together with the fitted scaling laws in the left 
panels of Supplementary Figure \ref{fig:scaling-relations}.

For optical-infrared reflectors (OIR) the large-sized and medium-sized categories were analysed together, 
with most of the medium-sized telescopes being of the ``monolithic thick mirror'' type, extending (and hence 
stabilising) the fit of the scaling relation towards smaller diameters.
Further telescopes types used were monolithic thin mirrors with active optics, segmented mirrors
that are mostly used for larger telescopes, and liquid mirrors for which a few instances exist in our
inventory.
A few telescopes were excluded from the fit of the construction cost -- diameter relation to avoid biasing 
by facilities that obviously do not follow the general trend.
The Vera Rubin Observatory (VRO) lies on the high side of the general trend of monolithic thin mirrors,
plausibly due to the complexity of the system and the camera.
The Hobby-Eberly Telescope (HET) and the Southern African Large Telescope (SALT), which is closely
based on the former, are well known to have low construction costs, typically 5--6 times less expensive
than other telescopes of comparable size.\cite{supp:leverington2017}
The ELT construction cost estimate also falls significantly below the trend line for ``segmented mirrors'', 
yet as construction is currently in progress the final cost is actually not yet known.
Furthermore, the ELT construction cost estimate is for the year 2023, which is a period with significant 
inflation, adding a further uncertainty to the estimate.
Our analysis suggests that monolithic thick mirrors are slightly more expensive compared to monolithic 
thin or segmented mirrors of the same size, an observation that was made already by 
ref.~\citen{supp:vanbelle2004}.
Power indices are found to be within $n\sim2.1-2.4$, which is compatible with literature values.

\begin{figure}
\centering
\includegraphics[width=15cm]{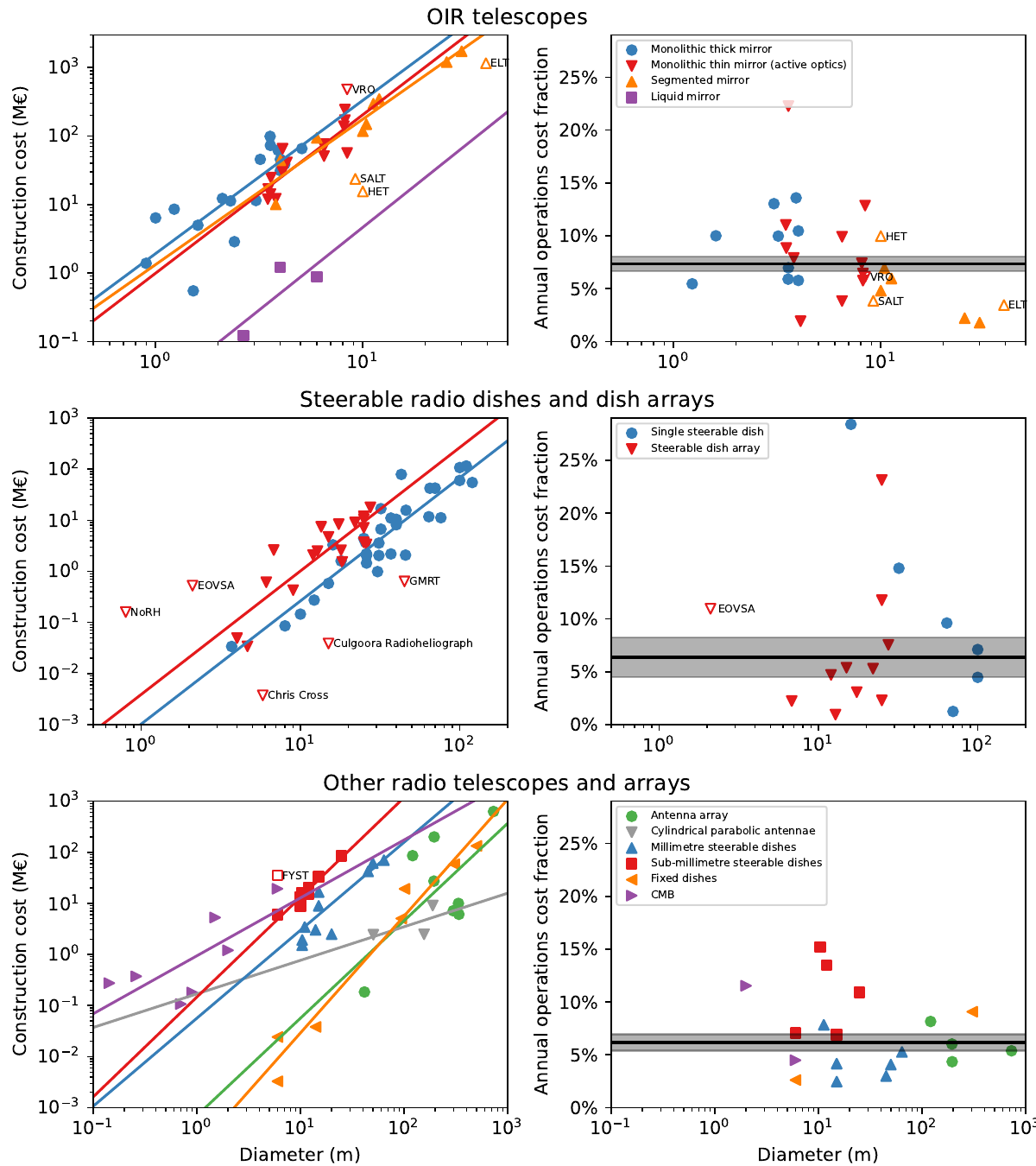}
\caption{
\small
{\bf Construction cost and annual operations cost fraction versus diameter for ground-based facilities.}
{\rm
Data are shown for OIR telescopes (top), steerable radio dishes and dish arrays (centre), and
other radio facilities (bottom), with different symbol colours used for different telescope types.
Left panels show construction costs, with symbols representing the data and lines showing the
best-fitting scaling relations.
Right panels show the fraction of annual operations cost over construction cost, with the
horizontal lines representing the mean value and the grey band its uncertainty that was
determined by successively excluding the five largest outliers from the mean values.
Data shown as open circles were excluded from the fit of the scaling relations.
}
\label{fig:scaling-relations}
}
\end{figure}

For radio telescopes we first considered single steerable dishes and dish arrays which are the telescope 
types for which we have most cost data.
For the telescope arrays, we used as $D$ the diameter of individual telescopes in the arrays, and as 
$C$ the total construction cost divided by the number $N$ of telescopes in the array, taking hence
the per-telescope cost for our analysis.
A few steerable dish arrays were excluded from the fits because they obviously present outliers with
respect to the trend observed for the other telescopes.
Those are
the Nobeyama Radioheliograph (NoRH),
the Expanded Owens Valley Solar Array (EOVSA),
the Chris cross, 
the Culgoora Radioheliograph,
and the Giant Metrewave Radio Telescope (GMRT).
Our analysis suggests similar power indices of $n\sim2.4$ for single steerable dishes and steerable
dish arrays, yet with a normalisation that indicates significantly larger per-telescopes costs for the 
arrays with respect to single dish observatories.
This may indicate that additional costs of combining individual antennae into an array may become
important for steerable dish arrays.

\begin{table}
\caption{{\bf Parameters of scaling relations for construction cost.}
{\rm
See Supplementary Equation \ref{eq:cost} for the meaning of $k$, $n$, and $D_0$.
}
\label{tab:scalingrelations}}
\centering
\vspace{6pt}
\begin{tabular}{l r c c}
\hline
Type & \multicolumn{1}{c}{$k$} & $n$ & Relative uncertainty \\
\hline
\multicolumn{4}{c}{\hrulefill OIR telescopes ($D_0$=5 m) \hrulefill} \\
Monolithic thick mirror & 70.50 \meuro\ & 2.238 & 70\% \\
Monolithic thin mirror & 41.21 \meuro\ & 2.314 & 38\% \\
Segmented mirror & 40.51 \meuro\ & 2.124 & 35\% \\
Liquid mirror & 0.871 \meuro\ & 2.415 & 69\% \\
\multicolumn{4}{c}{\hrulefill Radio dish(es) ($D_0$=10 m) \hrulefill} \\
Single steerable dish & 263.31 \keuro\ & 2.411 & 81\% \\
Steerable dish array & 1,016.00 \keuro\ & 2.417 & 91\% \\
\multicolumn{4}{c}{\hrulefill Other radio telescopes and arrays ($D_0$=10 m) \hrulefill} \\
Antenna array & 56.48 \keuro\ & 1.908 & 366\% \\
Cylindrical parabolic antennae & 766.97 \keuro\ & 0.657 & 43\% \\
Millimetre steerable dishes & 2,975.34 \keuro\ & 1.722 & 75\% \\
Sub-millimetre steerable dishes & 13,130.46 \keuro\ & 1.953 & 16\% \\
Fixed dishes & 29.05 \keuro\ & 2.287 & 81\% \\
CMB array & 12,796.54 \keuro\ & 1.134& 125\% \\
\hline
\end{tabular}
\end{table}

For all other radio telescope types we have less cost data, and eventually the diameters
span only a limited range, making it more difficult to define solid scaling relations.
Since our data suggest no obvious difference in the per-telescope cost for the other
radio telescope types, we combined single radio telescopes with radio telescope arrays for
our analysis.
Specifically, we distinguish 
antenna arrays (which are usually arrays of dipole antennae),
cylindrical parabolic antennae,
steerable millimetre and sub-millimetre dishes,
fixed dishes,
and telescopes for the observation of the cosmic microwave background (CMB).
For antenna arrays and cylindrical parabolic antennae, the collection area $A$ of the arrays 
were converted into effective diameters $D$ using $D=2\sqrt{A/\pi}$.
We consider the full collection area of the antenna arrays in our analysis, as the number
$N$ of antenna elements is not always well defined for these facilities.
A single facility was excluded from the determination of the scaling relations, which is the
Fred Young Submillimeter Telescope (FYST), which combines a novel design with special 
characteristics that likely drive the construction cost beyond the trend for other sub-millimetre 
telescopes.
Generally, the cost data show a larger scatter for the arrays compared to the single dish data,
suggesting that there are further underlying parameters that drive the cost, probably related
to combining the different antennas into an array.\cite{supp:boonstra2018}
For millimetre and sub-millimetre facilities we find power indices of $n\sim1.7$ and $n\sim2.0$ 
that are flatter than those of other steerable dishes, with significantly larger cost normalisations
$k$ that reflect the increasingly higher demand on surface roughness when moving to smaller
operational wavelengths.
Fixed dishes show a somewhat steeper power index of $n\sim2.3$, at a significantly smaller
cost with respect to steerable dishes, owing to the lack on a movable support structure.
Antenna arrays share the same area in the cost-diameter diagram than the fixed dishes,
yet with a large scatter.
Cylindric parabolic antennae show a distinctive trend with a much flatter power index, which we 
formally fitted to $n\sim0.7$, but which is only poorly constrained as only three facilities contribute to
the fit.
Finally we found a power index of $n\sim1.1$ for CMB facilities, though with a significant scatter, 
suggesting that there are further underlying parameters that drive the cost.

\begin{table}
\caption{{\bf Median annual operations cost fractions.}
\label{tab:oper-cost-fractions}}
\centering
\vspace{6pt}
\begin{tabular}{l r c c}
\hline
Category & Fraction & Uncertainty \\
\hline
OIR telescopes & 7.3\% & 0.7\% \\
Radio dish(es) & 6.4\% & 1.9\% \\
Other radio telescopes and arrays & 6.2\% & 0.8\% \\
\hline
\end{tabular}
\end{table}

Concerning operation costs, they are often considered to be a fraction of the construction costs
with typical values of 3--6\%, with a possible trend of decreasing fractions with increasing 
costs.\cite{supp:goodrich2019}
We show in the right panels of Supplementary Figure \ref{fig:scaling-relations} the annual operations 
cost fraction versus the collection diameter for the ground-based facilities in our inventory, which
primarily reveals an important scatter of individual data points.
Although the largest OIR telescopes show indeed the smallest operations cost fractions, this may
also be interpreted as a chance coincidence, as no obvious general trend of decreasing operations
cost fractions is visible in our data for any of the facility categories.
We therefore adjust the annual operations cost fractions by constants that are shown in Supplementary 
Figure \ref{fig:scaling-relations} as solid black lines.
By successively excluding the five largest outliers with respect to the constants we assess their
uncertainties, which are shown as grey bands in Supplementary Figure \ref{fig:scaling-relations}.
The results of this analysis are summarised in Supplementary Table \ref{tab:oper-cost-fractions}.
Our analysis suggests annual operations cost fractions of about 6--7\%, at the upper side of the values
found by ref.~\citen{supp:goodrich2019}.

\section*{Evolution of carbon intensity}

Emission factors of astronomical facilities are expected to decrease over the years due to reductions 
in the carbon intensity of energy production (for example as result of replacing fossil fuels by 
renewable energies) and due to improvements in the energy efficiency of construction processes 
and operation activities\cite{supp:mccann2022}.
Since quantitative estimates of these evolutions do not exist for astronomical facilities, 
we assume that the emission factors follow the general decarbonisation trends that are observed for 
the economy at large.
The latter are readily traced by the carbon intensity of gross domestic product (GDP) that we show
in Supplementary Figure \ref{fig:carbon-intensity} for the period 1935--2100.
We derived the carbon intensity data by dividing annual \co\ emissions through inflation corrected
GDP based on purchasing power parity (PPP) so that the results do not depend on economic 
conditions and relate to the purchase of physical goods.
\co\ emissions were computed by multiplying C emissions from ref.~\citen{supp:friedlingstein2022}
with 3.664.
GDP PPP data were taken from ref.~\citen{supp:owd2024} which are based on World Bank data from 1990 
onwards that were backwards extended based on the growth rates implied by the Maddison Project 
Database.

\begin{figure}[!t]
\centering
\includegraphics[width=15cm]{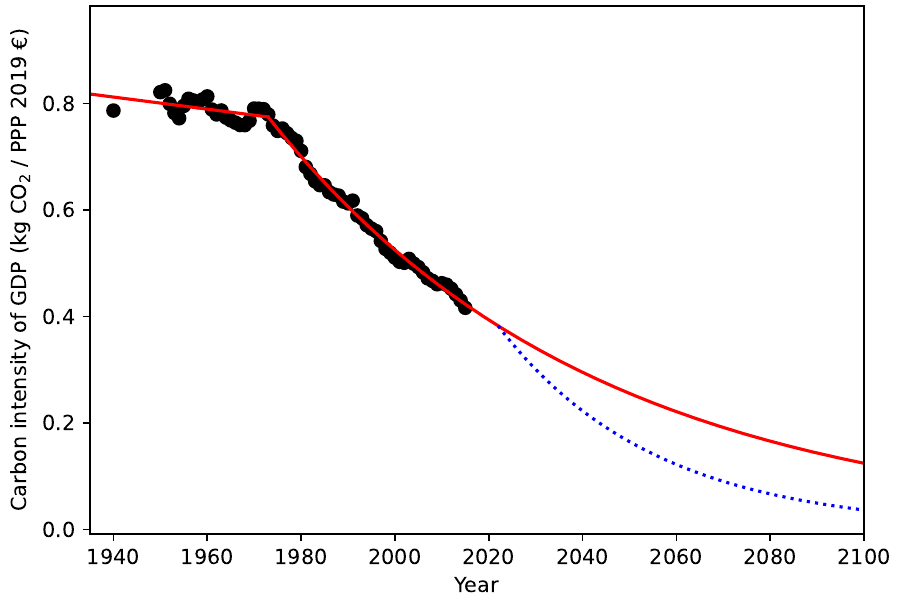}
\caption{
\small
{\bf Carbon intensity of gross domestic product (GDP).}
{\rm
Black dots show the observed carbon intensity of GDP.
The red solid line shows a piecewise function that was adjusted to the data.
The blue dotted line shows the carbon intensity evolution for the `Deep decarbonisation'
scenario.
}
\label{fig:carbon-intensity}
}
\end{figure}

We modelled the carbon intensity of GDP by a piecewise function composed of a linear
trend that is followed by an exponential decline.
We found that an adequate fit is obtained by setting the break of the function at the year
1973 which resulted in
\begin{equation}
CI_{\rm GDP}(t) = \left\{
\begin{array}{@{}ll@{}}
0.72 - 0.00113 \times (t-2019)& \mbox{for}\,\,t\le1973 \\
0.40 \times \exp\left(-0.0144 \times(t-2019)\right) & \mbox{else}
\end{array}
\right.
\label{eq:cigdp}
\end{equation}
where $t$ is time given in years and $CI_{\rm GDP}(t)$ is given in units of kgCO$_2$ per PPP 
2019 \euro.
Note that before 1973 the carbon intensity reduced by only about $1$ gCO$_2$ per PPP \euro\ and 
year, while after that year an exponential decrease with a rate of 1.44\% per year is observed.
Multiplying Supplementary Equation \ref{eq:cigdp} by 2.5 gives the relative carbon intensity
\begin{equation}
CI(t) = 2.5 \times CI_{\rm GDP}(t)
\end{equation}
that satisfies $CI(2019)=1$ and that we used in our work.

\section*{Illustration of annual carbon footprints}

To illustrate the modelling of the annual carbon footprint of astronomical facilities
we show in Supplementary Figure \ref{fig:annual-carbon-footprint} the models for a ground-based
observatory and a space mission.
The left panel shows the model for the 1 metre Schmidt telescope of the Stockholm observatory 
in Sweden that operated during the period 1963--2001.
As we could not determine the year of construction start for this telescope, the default construction 
duration of five years was adopted, which puts the construction start in the year 1958.
The model has an initial peak in the annual carbon footprint that corresponds to the
construction-related emissions that were distributed over the construction period.
The peak is followed by a tail corresponding to the operations-related emissions, which stop
at the moment when the operations cease.
For both contributions the annual carbon footprint decreases steadily with time due to the
decreasing carbon intensity of the emission factors, with an enhanced decrease from the year
1973 on that corresponds to the transition from a linear to an exponential decrease
(see Supplementary Figure \ref{fig:carbon-intensity}).

The right panel shows the model for the Hubble Space Telescope (HST), for which construction 
started in 1978 and that was launched in 1990.
The HST is still operating, hence the operations-related emission extend until the end year of 
2022 of our study.
Note that our model predicts that the construction-related emissions declined by about 15\% 
during the 12 years of construction due to the exponential decrease of the carbon intensity of 
the emission factors.
This decrease continues for the operations-related emission with a reduction by about 60\%
since the launch of the telescope.
 
\begin{figure}[!t]
\centering
\includegraphics[width=15cm]{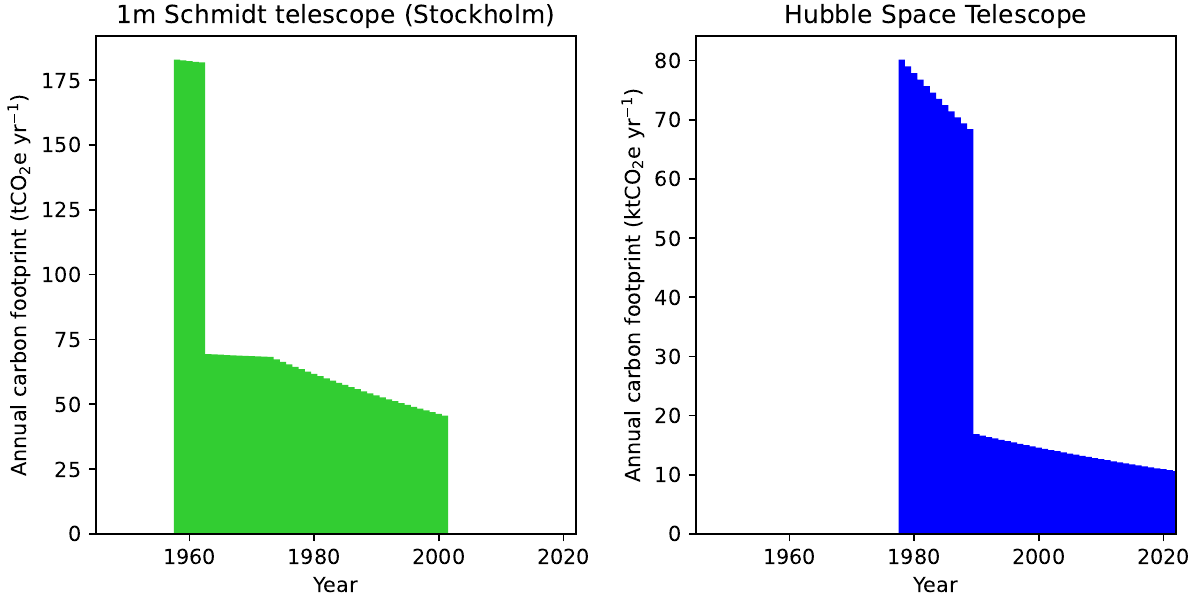}
\caption{
\small
{\bf Illustrations of annual carbon footprints.}
{\rm
The left panel shows the modelled annual carbon footprint of the 1 metre Schmidt telescope of the
Stockholm observatory, the right panel shows the modelled annual carbon footprint of the Hubble 
Space Telescope (HST).
Note that the annual carbon footprint of the 1 m Schmidt telescope is shown in units of \tcoeyr,
while that of the HST is shown in units of \ktcoeyr.
}
\label{fig:annual-carbon-footprint}
}
\end{figure}

\section*{Evolution of number of astronomers}

In order to understand how the evolution of the number of active astronomical facilities compares
to the evolution of the size of the community, we assessed the latter using the publication database
of the Astrophysics Data System (ADS).
Expanding on the method developed by ref.~\citen{supp:knoedlseder2022} we extracted from ADS for
each year of the period 1940--2022 the names of all authors that have authored publications in 
refereed journals assigned by ADS to the ``astronomy'' collection only.
Simplifying the names by retaining the surname and the initials of the first names, we determined
the list of individual authors for each year together with their affiliations.
From this we created the subset of individual authors that were affiliated to astronomical institutes 
that we retained as a measure of the size of the astronomical community.

The resulting evolution of the number of authors is shown in Supplementary Figure \ref{fig:authors}.
The number of affiliated authors shows a steady rise over the years, with a break in the growth rate
in the year 1975 and a pronounced dip in the year 1996.
Both features are linked to the way how ADS links affiliations to authors, illustrated by the fact that
the total number of authors shows no break in 1975, but rather a smooth decline for years earlier than 
1972, and does not reveal a dip in 1996.
Apparently, ADS affiliations of authors lack severe completeness before 1975 and had some issues
in the year 1996.
Comparing the number of affiliated US authors, shown as dotted brown line, to the number of members
of the American Astronomical Society (AAS)\cite{supp:abt2000}, shown as orange triangles, reveals 
furthermore that the drop in the number of authors before 1975 is probably related to the incompleteness 
of the ADS database, since the number of AAS members grows steadily without any change of slope.
We note that the number of AAS members is substantially larger than our assessment of the number
of affiliated US authors, which is understandable since AAS members also include physicists, mathematicians, 
geologists, engineers and others whose research interests lie within the broad spectrum of subjects now 
comprising contemporary astronomy.
To gauge whether the total size of the community is of the right order, we also show two literature
values\cite{supp:knoedlseder2022,supp:jascheck1991} that were derived independently, and which confirm 
our numbers. 

\begin{figure}[!t]
\centering
\includegraphics[width=15cm]{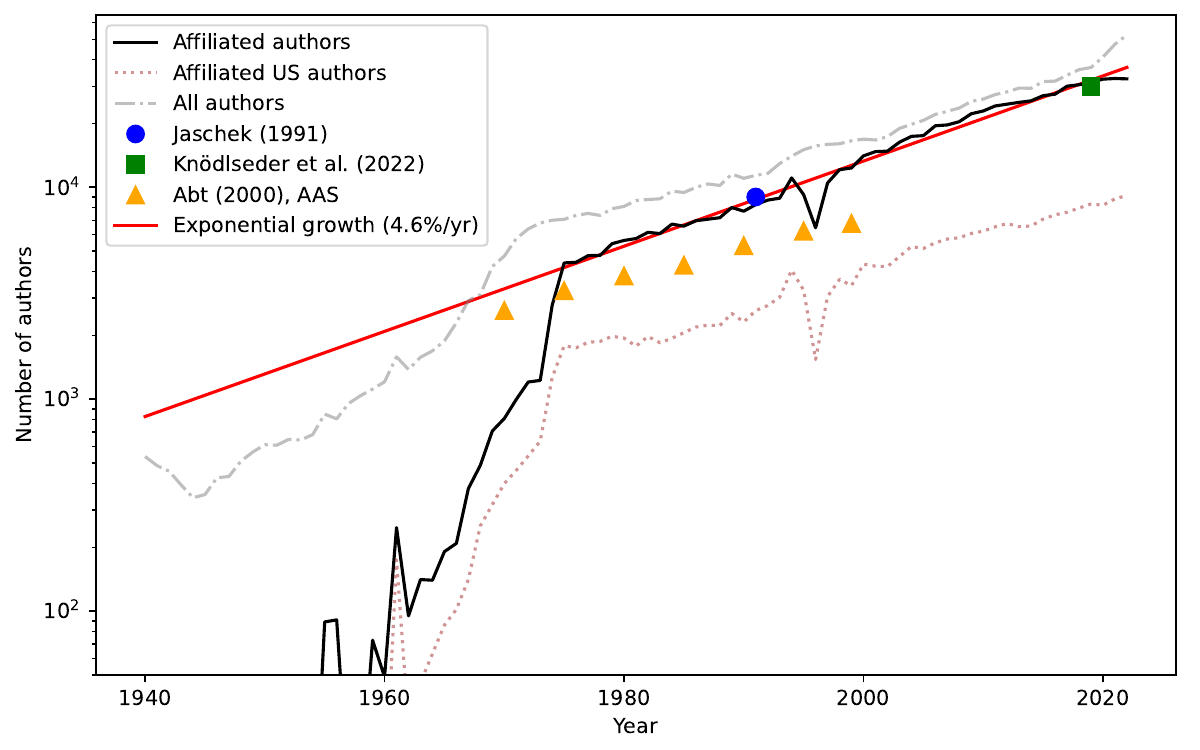}
\caption{
\small
{\bf Evolution of the number of affiliated authors.}
{\rm
The solid back line shows the evolution of the number of affiliated individual authors, the brown
dotted line shows the subset of US authors, and the grey dashed-dotted line shows all individual
authors in the ADS database.
Some quoted estimates of the community size are shown as symbols.
The red line shows an exponential growth model that was adjusted to the number of affiliated 
individual authors over the period 1975--2022.
}
\label{fig:authors}
}
\end{figure}

Since the growth of the astronomical community appears to be steady and exponential over the investigated
period, we fitted the number of affiliated authors by an exponential function, adjusted over years
1975--2022 to avoid years of incompleteness.
The fit suggests that the number of authors evolves like
\begin{equation}
\mbox{\it Number of individual authors}(t) = 32018 \times \exp\left(0.046 \times(t-2019)\right)
\label{eq:authors}
\end{equation}
with an exponential growth rate of 4.6\% per year.
This corresponds to a doubling time of 15 years of the astronomical community, which corresponds
to the doubling time that was already suggested in the 1960s for the growth of the scientific 
community.\cite{supp:desollaprice1963}
The fitted function is shown as red line in Supplementary Figure \ref{fig:authors}, illustrating that the
function is a reliable representation of the data.

\section*{Evolution of average facility lifetimes}

To investigate the origin of the growth in the number of operating facilities, we determined the
average lifetime of all facilities that were operating in a given year and show the result over
the period of the study in Supplementary Figure \ref{fig:lifetimes}.
Only facilities were considered for which start and end dates for operations were available in our
inventory.
Initially, ground-based facilities had a rather long lifetime of over 60 years, yet this number dropped
subsequently to settle in a band covering 28--42 years, with a typical lifetime of 38 years.
Remarkably, the average lifetime of facilities that were operating around the year 2013 dropped to
the lowest average lifetime of 28 years, which we tracked back to a drop in lifetimes of radio telescopes, 
yet longer facility lifetimes were quickly recovered a few years later.
For space missions the dynamic is opposite, with a gradual increase of facility lifetimes from one
year (which is the resolution of our analysis) to a typical value of 18 years in about 1990, staying in
a narrow band from 14--20 years.
From 2010 the average facility lifetime starts to drop, which we tracked back to the contribution
of nano satellites that have usually a relatively short lifetime.
As neither ground-based facilities nor space missions show an increasing trend in the average facility
lifetime over the last decades, we attribute the observed increase in the number of operating facilities
primarily to an increase in the deployment rate of new facilities.

\begin{figure}[!t]
\centering
\includegraphics[width=15cm]{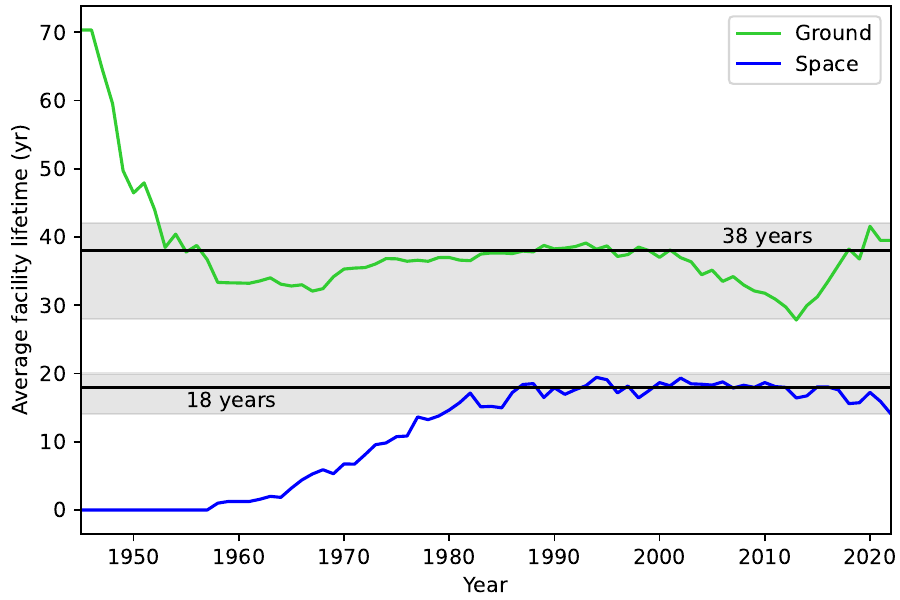}
\caption{
\small
{\bf Evolution of the average facility lifetime.}
{\rm
The green and blue curves show the evolution of the average lifetime of operating ground-based
and space-based facilities, respectively.
The minimum and maximum variation of the average facility lifetime over the last decades are indicated 
by grey bands, typical values are shown as solid black lines.
}
\label{fig:lifetimes}
}
\end{figure}

\section*{Evolution of planetary exploration missions}

\begin{figure}[!t]
\centering
\includegraphics[width=16.5cm]{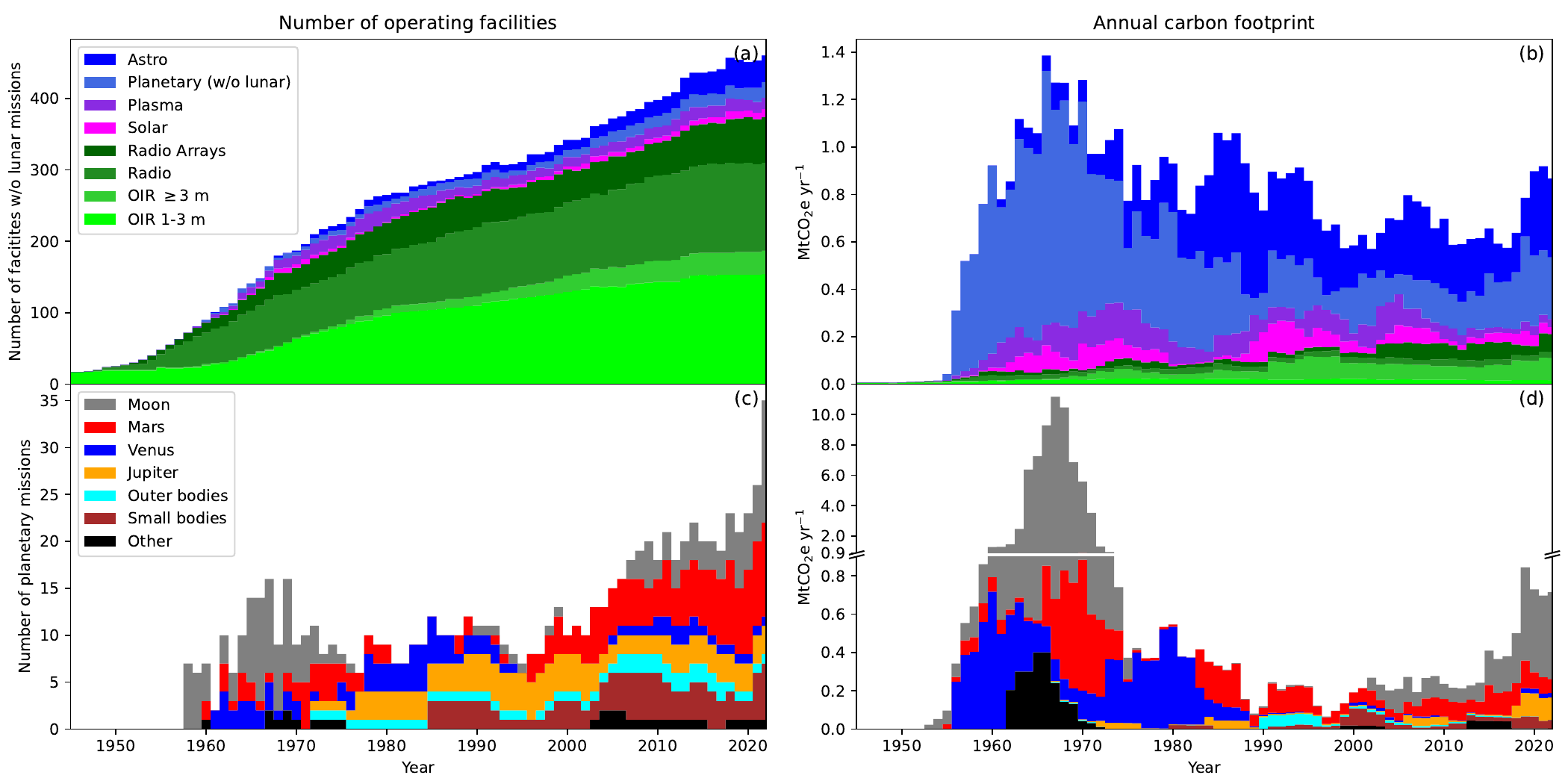}
\caption{
\small
{\bf Evolution of the number of operating facilities and their carbon footprint excluding lunar
exploration missions, and breakdown of planetary missions by explored solar system body.}
{\rm
Panels (a) and (b) show the evolution of the number of operating facilities and their annual carbon
footprint when lunar exploration missions are excluded.
Panels (c) and (d) show the evolution of the number of operating planetary exploration missions and
their annual carbon footprint, split by explored solar system body.
Outer bodies include Saturn, Uranus, Neptun and Pluto.
Small bodies include missions to comets, asteroids and planetary defence missions.
Other missions include test missions or missions with unknown mission objective.
}
\label{fig:evolution-planetary}
}
\end{figure}

Our analysis revealed that planetary exploration missions were the primary driver of the GHG 
emissions at the beginning of the space age, and even today they still account for about half of
the annual carbon footprint of astronomical facilities.
As lunar exploration was a primary driver for this footprint in the 1960--1970s, and since one may
argue that the motivation of the lunar exploration missions was primarily geopolitical and not 
scientific, we show in panels (a) and (b) of Supplementary Figure \ref{fig:evolution-planetary} the 
evolution of the number of astronomical facilities and their carbon footprint without taking into 
account any lunar missions.
While this only removes a small hump in the number of operating facilities, the peak in the annual 
carbon footprint of 11.8 \Mtcoeyr\ at the end of the 1960s is is now drastically reduced to
1.4 \Mtcoeyr.
Yet there is still a peak of GHG emissions in this early period of the space age, and panels (c) and 
(d) of Supplementary Figure \ref{fig:evolution-planetary} reveal that this peak originates primarily from 
planetary exploration missions to Venus and Mars.
Missions to other planets contribute only at a low level to the annual carbon footprint over the
years, although the construction of the Juno and JUICE missions to Jupiter become discernible at 
the end of the 2000s and around 2020, owing to the relatively large payload masses of 3.6 and 6.1 
tons, respectively.
Interestingly, while lunar missions have essentially ceased by the mid 1970s, their is renewed
interest in the exploration of the Moon that started in the mid 2000s, and that since the mid 2010s
dominates again the annual carbon footprint of planetary exploration.
Given the huge environmental impact of the first space race, there exists a real risk that the 
`NewSpace' race will be equally harmful.

\section*{Interpretation of the mean annual facility mass or cost}

\begin{figure}[!t]
\centering
\includegraphics[width=16.5cm]{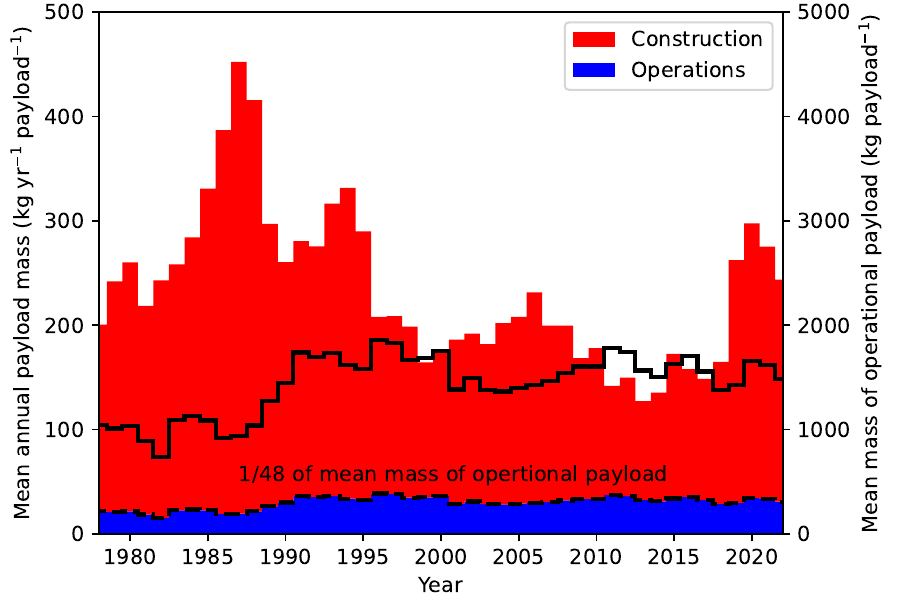}
\caption{
\small
{\bf Mean annual payload mass compared to the mean mass of operating payloads for the last 45 
years.}
{\rm
The mean annual payload mass is shown as histograms and the left scale, divided into contributions 
from construction (red) and operations (blue).
For comparison, the mean mass of operating payloads in space is shown as solid line and the right
scale.
The curve is also shown scaled by a factor of 1/48 as dashed line, now applying to the left scale.
}
\label{fig:payload-mass}
}
\end{figure}

The mean annual payload mass is used in the carbon footprint factorisation as activity data, and 
measures the amount of GHG releasing activities in the space sector per facility that is already 
operating in space. 
As illustrated in Supplementary Figure \ref{fig:payload-mass}, most of these activities are related
to construction of space missions (red histogram), while operations contribute only little (blue 
histogram).
Since there is no direct causal link between construction activities on ground, and the number of 
already operating payloads in space, the mean annual payload mass does not follow the mean mass 
of payloads that are operating in space (black line).
Construction activities on ground will however impact the payload mass in space at later times,
as payload mass constructed on ground will add after launch to the payload mass in space.
This is clearly seen in the years 1985--1990, where the peak in the mean annual payload mass is
partly due to the contemporaneous construction of two heavy satellites, the Hubble Space Telescope 
(11.1 tons) and the Compton Gamma-Ray Observatory (16.3 tons).
Following the launch of both missions in the years 1990 and 1991, respectively, the mean annual payload 
mass drops while the mean mass of operating payloads rises.
In contrast, operation activities are directly linked to the payload mass in space, as illustrated by the
perfect match between the blue histogram and the dashed curve in Supplementary Figure 
\ref{fig:payload-mass}.
Consequently, this contribution will depend directly on the mean mass of operating payloads,
which since 1990 is rather stable, with a typical value of 1.5 tons per payload.
The mean annual payload mass is thus expected to show a sawtooth pattern on top of a relatively
flat profile, with peaks corresponding to periods of intense construction of heavy payloads, followed 
by decays once these missions are launched into space.

\begin{figure}[!t]
\centering
\includegraphics[width=16.5cm]{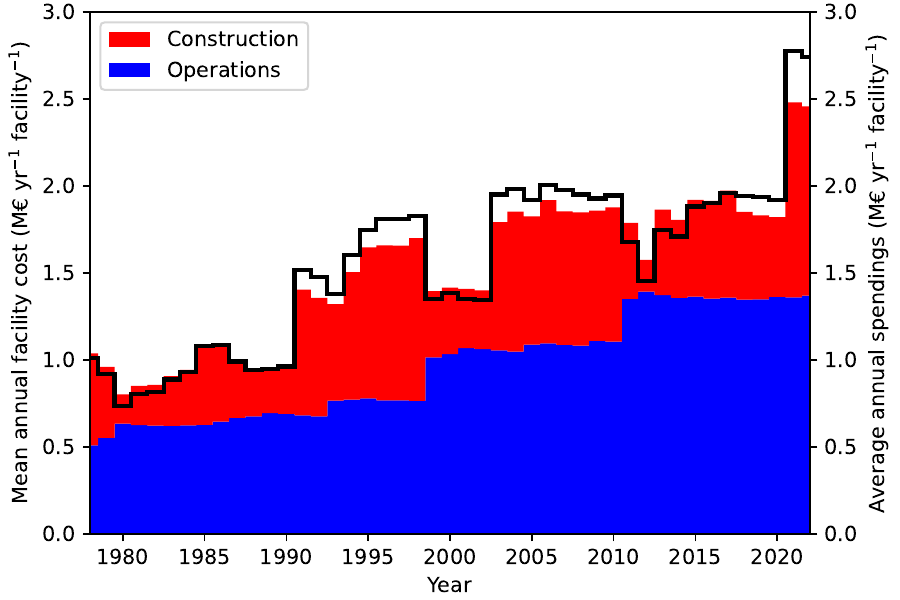}
\caption{
\small
{\bf Mean annual facility cost compared to the annual spendings per facility for the last 45 years.}
{\rm
The mean annual facility cost is shown as histograms and the left scale, divided into contributions 
from construction (red) and operations (blue).
For comparison, the average annual spendings on ground-based observatories are shown as solid 
line and the right scale.
}
\label{fig:facility-cost}
}
\end{figure}

The mean annual facility cost for ground-based observatories is analogous to the mean annual payload 
mass, and measures the amount of GHG releasing activities per operating observatory. 
Supplementary Figure \ref{fig:facility-cost} shows the mean annual facility cost separated into
contributions from construction and operations.
Owing to the comparable emission factors for construction and operations of ground-based observatories
combined with an average operating life time that is superior to the typical construction period, operations 
dominate the mean annual facility cost for ground-based observatories.
In comparing the mean annual facility cost to the average annual spendings on ground-based 
observatories a delay between construction spendings and the mean annual facility cost is observed 
that is equivalent to the situation for space missions.
Since the number of operating facilities does not include facilities under construction, the average
annual spendings exceed the mean annual facility cost, but drop back to a comparable level once
the newly constructed observatories enter operations.

\section*{Different modellings of carbon footprint factorisation coefficients}

\begin{figure}[!t]
\centering
\includegraphics[width=16.5cm]{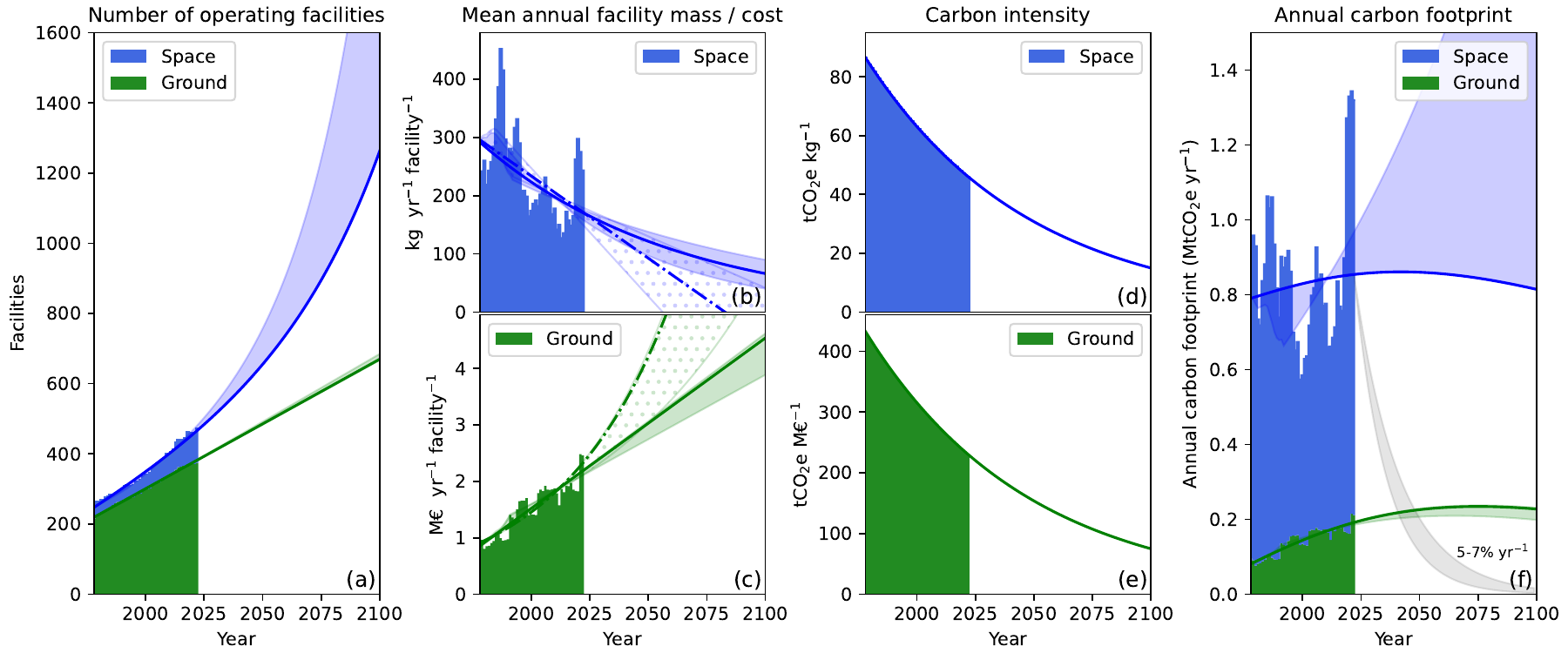}
\caption{
\small
{\bf Reference modelling.}
{\rm
The panels (a--f) show the carbon footprint factorisation when using an exponential law to model
the number of operating space missions and a linear law to model the number of operating 
ground-based facilities.
Shaded bands were obtained by varying the adjustment period of the analytical laws from the last 
30 to 45 years. 
Lines corresponding to adjustments over the last 45 years.
The grey shaded band in panel (f) corresponds to the annual emission reduction target that is
comprised with 5-7\%.
}
\label{fig:modelling-mix}
}
\end{figure}

The panels (a--d) of Figure \ref{fig:scenarios} present different modellings of the carbon footprint 
factorisation coefficients, and we show in Supplementary Figures \ref{fig:modelling-mix}--\ref{fig:modelling-payload} 
the corresponding carbon footprint factorisations and the adjusted analytical functions.
Supplementary Figure \ref{fig:modelling-mix} presents the reference modelling, where we adjusted the 
number of operating space missions using an exponential law while for the number of operating 
ground-based facilities we used a linear law.
For the mean annual facility mass of space missions we used an exponential law, while for illustrative 
purposes we also show a linear law.
For the mean annual facility costs of ground-based observatories we did the opposite, using a linear
law for the analysis and showing an exponential law for illustrative purposes.
The carbon intensities were adjusted by exponential laws for space and ground-based facilities. 
We used the same laws for the mean annual facility mass or cost and the carbon intensities in 
Supplementary Figures \ref{fig:modelling-exp}--\ref{fig:modelling-payload}.
The blue curve and band corresponds to the `Research as usual' scenario, shown as red curve and
band red in panel (a) of Figure \ref{fig:scenarios}.

\begin{figure}[!t]
\centering
\includegraphics[width=16.5cm]{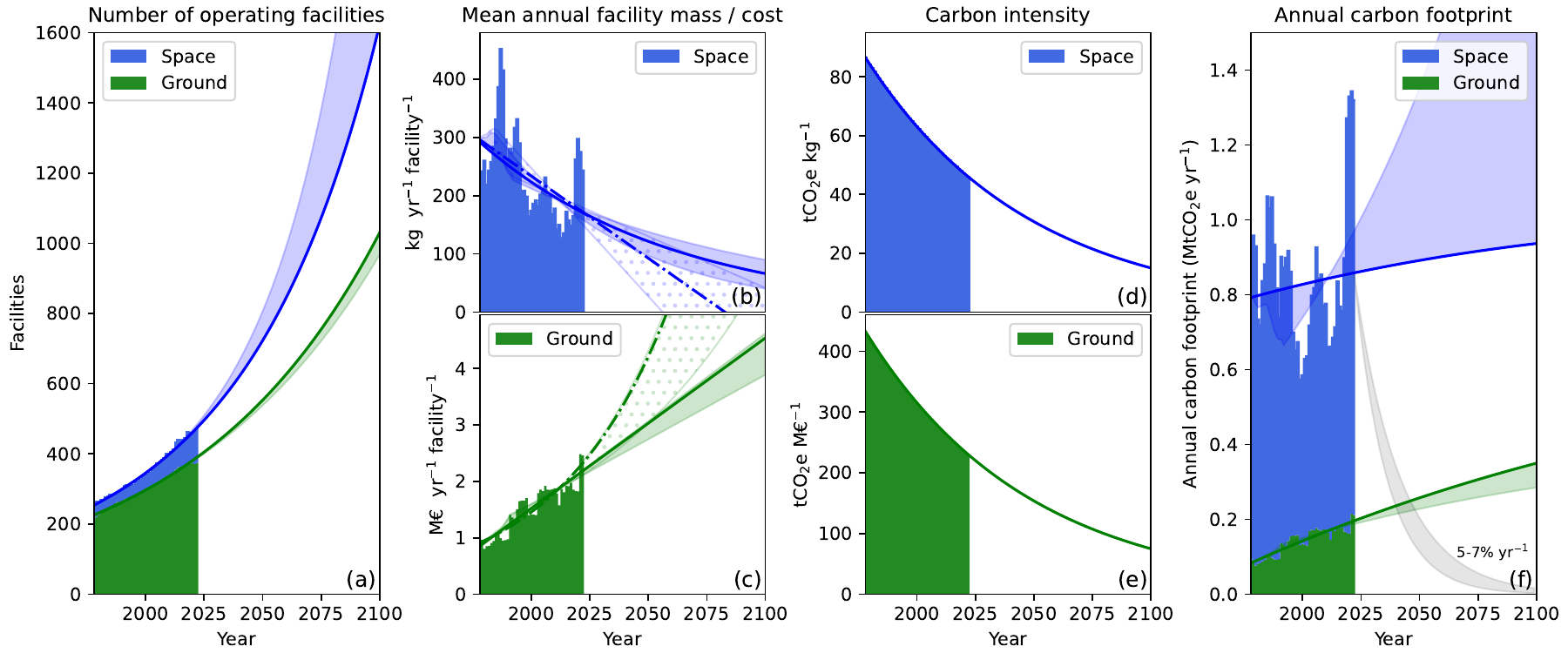}
\caption{
\small
{\bf Exponential laws for number of operating facilities.}
{\rm
Carbon footprint factorisation using exponential laws to model the number of operating space missions.
Shaded bands were obtained by varying the adjustment period of the analytical laws from the last 
30 to 45 years. 
Lines corresponding to adjustments over the last 45 years.
The grey shaded band in panel (f) corresponds to the annual emission reduction target that is
comprised with 5-7\%.
}
\label{fig:modelling-exp}
}
\end{figure}

In Supplementary Figure \ref{fig:modelling-exp} we adjusted the number of operating space missions 
and ground-based observatories by exponential laws, which leads to an increased prediction of
the annual carbon footprint with respect to Supplementary Figure \ref{fig:modelling-mix} as a result 
of an enhanced contribution from ground-based observatories.

\begin{figure}[!t]
\centering
\includegraphics[width=16.5cm]{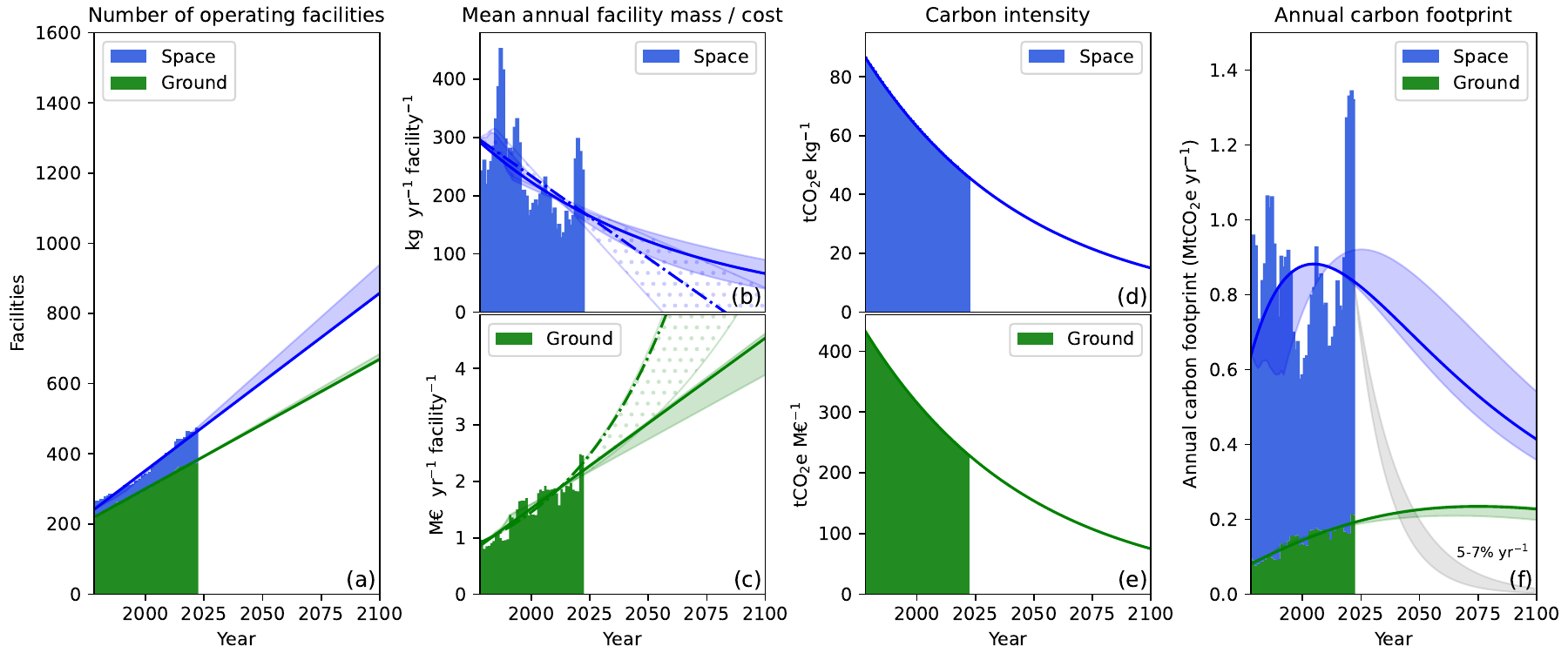}
\caption{
\small
{\bf Linear laws for number of operating facilities.}
{\rm
Carbon footprint factorisation using linear laws to model the number of operating ground-based and 
space facilities.
Shaded bands were obtained by varying the adjustment period of the analytical laws from the last 
30 to 45 years. 
Lines corresponding to adjustments over the last 45 years.
The grey shaded band in panel (f) corresponds to the annual emission reduction target that is
comprised with 5-7\%.
}
\label{fig:modelling-lin}
}
\end{figure}

In Supplementary Figure \ref{fig:modelling-lin} we replaced both exponential laws for the number of 
operating facilities by linear laws, leading to a much reduced prediction in the number of operating 
facilities by the end of the century.
With these laws the annual carbon footprint is showing a curved shape, owing to the exponential
decrease of the mean annual facility mass for space missions and the carbon intensities that
overtakes the rise in the number of operating facilities.
Nevertheless, the resulting reduction in the annual carbon footprint is still well above the required
suggested carbon footprint reduction trajectory for astronomical facilities.

\begin{figure}[!t]
\centering
\includegraphics[width=16.5cm]{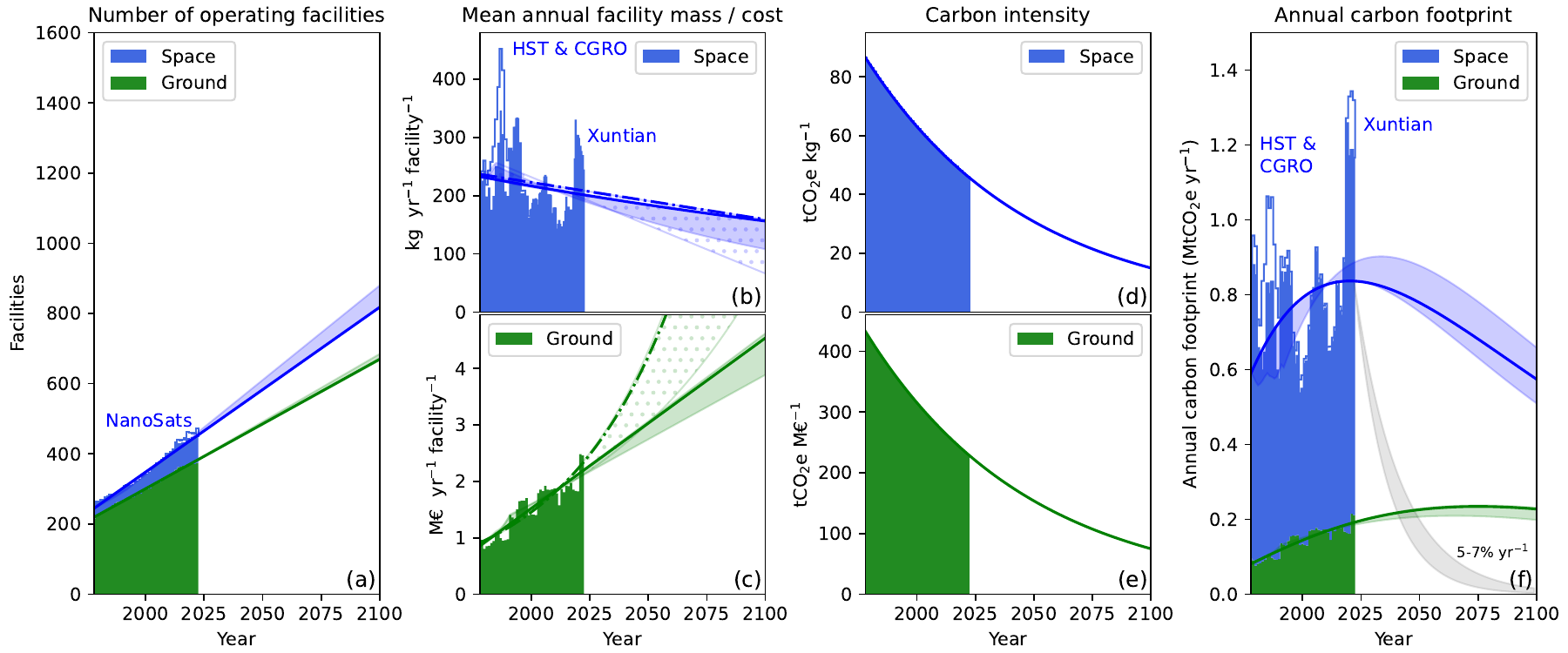}
\caption{
\small
{\bf Excluding payloads with small and large masses.}
{\rm
Carbon footprint factorisation using linear laws to model the number of operating ground-based and 
space facilities.
Payloads with masses of $<$20 kg (NanoSats) and $>$10 tons (HST, GRO \& Xuntian) 
were excluded from the data.
Shaded bands were obtained by varying the adjustment period of the analytical laws from the last 
30 to 45 years. 
Lines corresponding to adjustments over the last 45 years.
The grey shaded band in panel (f) corresponds to the annual emission reduction target that is
comprised with 5-7\%.
}
\label{fig:modelling-payload}
}
\end{figure}

In Supplementary Figure \ref{fig:modelling-payload} we show the same modelling of the number of 
operating facilities by linear laws, but now we excluded all payloads with masses smaller than 
$<$20 kg and larger than $>$10 tons from the adjustments.
The lower mass limit is motivated by the rising contribution of nano satellites (NanoSats) to the
global inventory of operating space missions, that drive the number of operating facilities upwards
and the mean annual facility mass for space missions downwards, while contributing little to the
carbon footprint as result of their small mass.
In total, 25 NanoSats were excluded by applying the $<$20 kg mass limit.
The upper mass limit is motivated by the large contribution of heavy satellites to the mean annual 
facility mass for space missions, as seen by the important spikes in the panels (b) of the figures.
Applying an upper mass limit of 10 tons excludes the Hubble Space Telescope (HST), the
Compton Gamma-Ray Observatory (CGRO) as well as the Xuntian space telescope, which is currently
under construction in China, from the adjustment.
In particular, the HST and CGRO were constructed contemporaneously, hence their construction emissions
pile up, leading to an important peak in the mean annual facility mass and the carbon footprint for
the years 1985--1990.
Applying the mass limits results in fact in a much flatter fitted trend for the mean annual facility mass 
distribution for space missions that is very similar for exponential and linear laws.
This indicates that the exponential decay that is fitted without the mass limits is largely produced by
the NanoSats and a few heavy payloads, and that the actual decay when these payloads are
excluded is less pronounced.
A rather flat trend is in fact expected due to the relatively constant mean mass of operating payloads
over the last decades, as revealed in Supplementary Figure \ref{fig:payload-mass}, although we recall
that the mean annual facility mass is primarily determined by construction activities, and it may well
be that astrophysical payloads that are currently under construction have on average less mass than 
those that are already in orbit.

\section*{Using mission cost instead of payload launch mass as activity data for space missions}

In Paper I, two options were discussed to model the carbon footprint of space missions, one using 
total mission cost and another using payload launch mass as activity data.
Since payload launch masses are readily available for all space missions and not subject to significant
uncertainties, we used them as activity data in our study.
To investigate the impact of this choice, we perform here an alternative modelling where mission cost
is used instead of payload launch masses as activity data.
Since we have no mission cost estimates for all space missions in our inventory, we estimate
missing data from a scaling law that we adjusted to our cost data, with mission cost varying in 
proportion to some power of the payload launch mass.
The results are shown in Supplementary Figure \ref{fig:correlation-space-cost}, resulting in
an almost linear relation
\begin{equation}
\mbox{Mission cost (\keuro)} = 297 \times \mbox{Payload mass (kg)}^{0.96}
\end{equation}
though with a considerable scatter (the mean absolute deviation of the mission cost from the
scaling law amounts to 84\%).

\begin{figure}[!t]
\centering
\includegraphics[width=15cm]{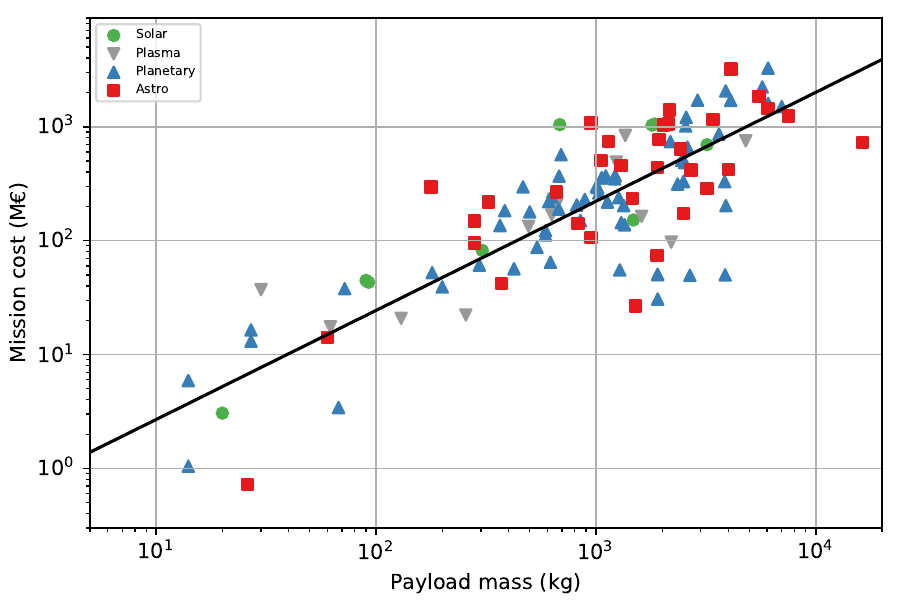}
\caption{
\small
{\bf Mission cost versus payload launch mass for space missions.}
\label{fig:correlation-space-cost}
}
\end{figure}

Analogous to the emission factors for payload mass, we derived from ref.~\citen{supp:wilson2023} emission
factors of 135 \monetaryef\ for construction and 2 \monetaryefyr\ for the operations of a space
mission, replacing Equation \ref{eq:space-footprint} for computing the carbon footprint by
\begin{equation}
F(t) = CI(t) \times \left\{
\begin{array}{@{}ll@{}}
135 \times \frac{\mbox{Mission cost (\meuro)}}{\mbox{$t_o -t_c$}} & \mbox{for}\,\,t_c \le t < t_o \\
2 \times \mbox{Mission cost (\meuro)} & \mbox{for}\,\,t_o \le t \le t_e \\
0 & \mbox{else}
\end{array}
\right.
\label{eq:space-cost-footprint}
\end{equation}
and replacing Equation \ref{eq:space-payload} by the annual space mission cost, given by
\begin{equation}
C_s'(t) = \sum_i \left\{
\begin{array}{@{}ll@{}}
\frac{\mbox{Cost of mission $i$ (\meuro)}}{\mbox{$t_o^i -t_c^i$}} & \mbox{for}\,\,t_c^i \le t < t_o^i \\
\frac{2}{135} \times \mbox{Cost of mission $i$ (\meuro)} & \mbox{for}\,\,t_o^i \le t \le t_e^i \\
0 & \mbox{else}
\end{array}
\right.
\label{eq:space-cost}
\end{equation}
The resulting modelling of the future annual carbon footprint of astronomical facilities, that can be 
directly compared to Figure \ref{fig:projection}, is shown in Supplementary Figure 
\ref{fig:modelling-space-cost}.
The predicted trends in the annual carbon footprint are similar to those obtained when using the
payload launch mass, though the absolute level of GHG emissions is reduced by about 25\%.

\begin{figure}[!t]
\centering
\includegraphics[width=16.5cm]{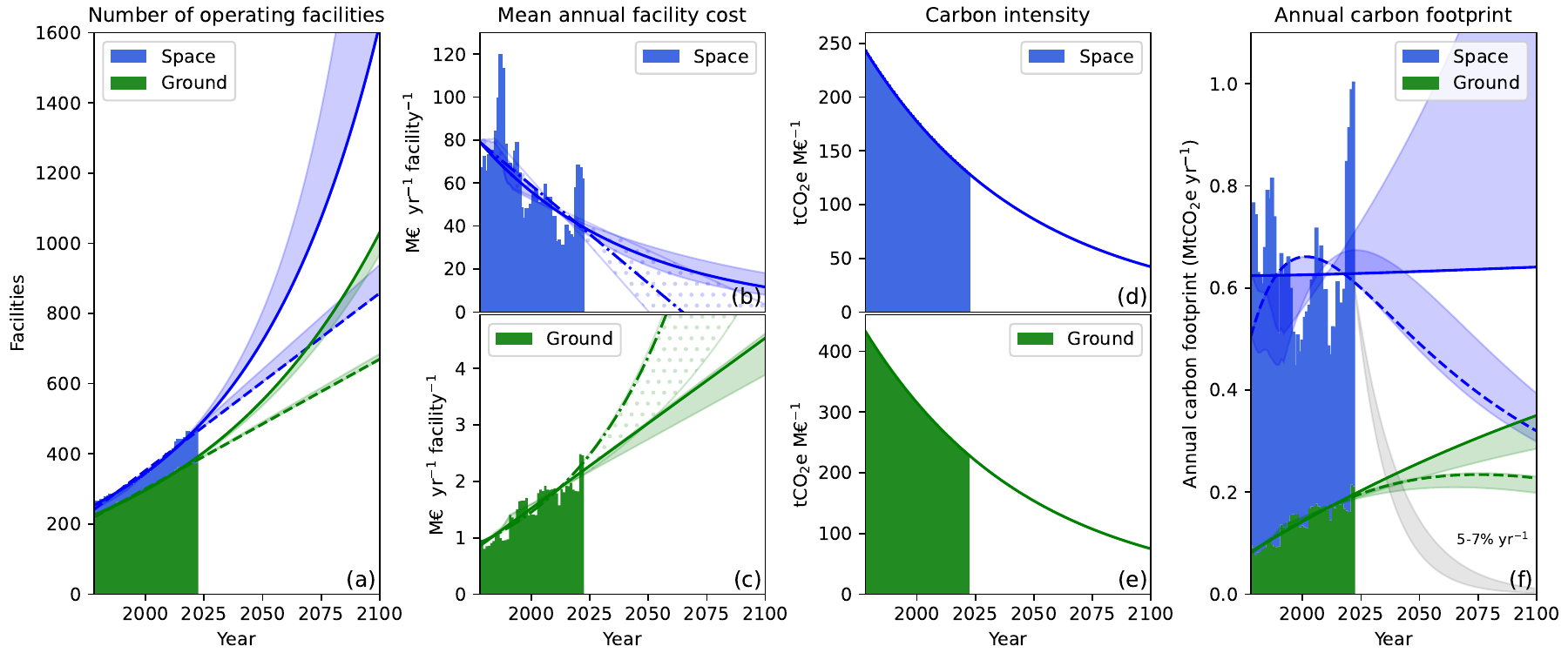}
\caption{
\small
{\bf Modelling of the future annual carbon footprint of astronomical facilities using mission cost instead 
of payload launch mass as activity data for space missions.}
{\rm
Shaded bands were obtained by varying the adjustment period of the analytical laws from the last 
30 to 45 years. 
Lines corresponding to adjustments over the last 45 years.
The grey shaded band in panel (f) corresponds to the annual emission reduction target that is
comprised with 5-7\%.
}
\label{fig:modelling-space-cost}
}
\end{figure}

\section*{Comparison to estimates of Kn\"odlseder et al. (2022)}

We compare in this section the annual carbon footprint results for the year 2019 obtained by 
Kn\"odlseder et al. (2022)\cite{supp:knoedlseder2022} (Paper I) to this work.
Paper I derived the emissions of the world-fleet of astronomical facilities by extrapolation 
from a list of 85 facilities that were used by astronomers at the Institut de Recherche en Astrophysique 
et Plan\'etologie (IRAP) in the year 2019.
In this work we avoided any extrapolation, and directly computed the annual carbon footprint for a 
complete census of 1,211 ground-based observatories and space missions.
In addition, while in Paper I, emissions from facility construction were released over the entire operating 
life time of each facility, and counted on top of emissions from operations, we used in this work a scheme 
where construction emissions were released during the construction period and emissions from 
operations during the operating life time.
This reduces the emissions for operating facilities in 2019, as construction emissions were already 
released before, but construction emissions generated from building facilities in 2019 are now added 
to the annual carbon footprint of this year.

\begin{table}
\caption{{\bf Comparison to estimates of Kn\"odlseder et al. (2022).}
{\rm
Carbon footprints are given in units of \ktcoeyr, columns marked by \# indicate number of facilities.
The number of facilities of `Paper I' are estimations of the world-fleet of facilities operating in the
year 2019.
The number of facilities of `This work' give the number of facilities in the inventory of this work
that were constructed or operating in the year 2019.
}
\label{tab:comparison}}
\centering
\vspace{6pt}
\begin{tabular}{l |r r|r r| r r r r}
\hline
& \multicolumn{2}{c|}{Paper I} & \multicolumn{6}{c}{This work}\\
& \multicolumn{2}{c|}{} & \multicolumn{2}{c|}{Total} & \multicolumn{2}{c}{Construction} & \multicolumn{2}{c}{Operations}  \\
Category & Footprint & \# & Footprint & \# & Footprint& \# & Footprint & \# \\
\hline
Solar & 23$\pm$16 & 3 & 60.8 & 19 & 52.7 & 9 & 8.1 & 10 \\
Plasma & 58$\pm$20 & 13 & 40.0 & 30 & 22.1 & 10 & 17.9 & 20 \\
Planetary & 226$\pm$54 & 21 & 844.4 & 57 & 803.8 & 36 & 40.6 & 21 \\
Astro & 149$\pm$43 & 18 & 163.2 & 65 & 104.3 & 24 & 59.0 & 41 \\
Total space & 455$\pm$74 & 55 & 1,108.4 & 171 & 982.9 & 79 & 125.6 & 92 \\
\hline
OIR ($\ge$3m) & 171$\pm$27 & 37 & 81.6 & 43 & 32.5 & 4 & 49.1 & 39 \\
OIR (others) & 147$\pm$6 & 1,000 & 16.7 & 174 & 1.4 & 4 & 15.3 & 170 \\
Radio & 127$\pm$18 & 74 & 23.1 & 131 & 4.3 & 5 & 18.8 & 126 \\
Radio arrays & 260$\pm$115 & 27 & 42.1 & 68 & 5.0 & 5 & 37.1 & 63  \\
Others & 52$\pm$53 & 4 & - & - & - & - & - & - \\
Total ground & 757$\pm$131 & 1,142 & 163.5 & 416 & 43.2 & 18 & 120.3 & 398 \\
\hline
Total & 1,212 & 1,197 & 1,271.9 & 552 & 1,026.1 & 92 & 245.9 & 460 \\
\hline
\end{tabular}
\end{table}

The estimates of the annual carbon footprint obtained in Paper I are compared in Supplementary Table 
\ref{tab:comparison} to the results obtained in this work per facility category (see also Table 4 of Paper I).
For space missions, the most notably difference is found for planetary exploration missions, where in 
addition to the 21 operating facilities that were estimated in Paper I there are 35 facilities under 
construction in 2019, which contribute a footprint of 804 \ktcoeyr\ that largely explains the observed 
difference.
Planetary exploration missions is in fact the only category where in 2019 more facilities are
under construction than operating, which is explained by the recent surge for such missions in the
context of the `NewSpace' race towards the Moon.
For all other categories, operating facilities dominate in number.
For other space mission categories, we also note that there are more operating space missions in the 
lists of this work compared to estimations made in Paper I, which is primarily due to small or nano satellites 
that were not considered in 
Paper I, and that will only moderately change the estimates of the annual carbon footprint due to
their small mass.

For ground-based observatories, our annual carbon footprint estimates fall systematically below those 
found in Paper I.
The number of ground-based observatories in the list of Paper I was actually very limited, and it turned 
out that this introduced a substantial bias towards too large carbon footprints in the extrapolation procedure, 
which becomes most visible for the `OIR (others)' category.
The bias is related to the fact that telescope costs, which we use as activity data for the carbon footprint 
estimates, scale with a power $n$ of the diameter, with powers of 2.1--2.4 found for example for OIR
telescopes (see Supplementary Table \ref{tab:scalingrelations}).
The list of telescopes that were considered in the `OIR (others)' category in Paper I comprised 19
operating telescopes of which 45\% had diameters larger than 2 m.
Our more extended list of 170 operating telescopes, that represents the complete world-fleet of 1--3 m OIR 
telescopes, contains only 13\% of telescopes with diameters larger than 2 m, revealing that the list 
considered in Paper I was biased towards telescopes with large diameters.
This is a natural consequence of the selection procedure that was used for the list of telescopes in Paper I, 
since astronomers tend to use large telescopes for their scientific publications due to their better performance.
Paper I extrapolated the proportion of 45\% telescopes with diameters larger than 2~m to a world-fleet 
of about 1,000 telescopes, predicting hence a total of 450 such telescopes that were operating in 2019 in 
the world.
Our list suggests, however, that only 22 such telescopes were operational in 2019, which implies an 
over-estimation by a factor of 20 of the number of telescopes in this category.
Similar, yet less extreme, situations exist for the other categories.
For example, in the `OIR ($\ge3$ m)' category, Paper I had 67\% telescopes with diameters of at least 8 m
in the list, while in our work the proportion is 36\%, hence an over-estimation of almost a factor of two.

Furthermore, from a total of 416 ground-based telescopes, only 18 (4\%) contributed emissions to 
construction in 2019 while the remaining emissions were attributed to operations, which is related to
the fact that ground-based observatories have a relatively long life time compared to their construction 
duration (see Supplementary Figure \ref{fig:lifetimes}).
Construction emissions hence fluctuate over the years, and for a particular year, they depend on the specific
construction activities.
To avoid such fluctuations, construction emissions in Paper I were distributed over the full operating
life times.
If we had adopted the same scheme in this work, the total emissions from `Radio arrays' would, for 
example, have increased from 42 \ktcoeyr\ to 113 \ktcoeyr, and the total 2019 emissions from ground-based 
facilities would have increased from 163 \ktcoeyr\ to 249 \ktcoeyr.
The emissions from space missions would also have been affected, in particular for planetary exploration 
mission, as discussed above, which would have dropped from 844 \ktcoeyr\ to 258 \ktcoeyr.
This would have changed the total emissions for space missions from 1,109 \ktcoeyr\ to 574 \ktcoeyr,
and the total GHG emissions of 2019 would have dropped from 1,272 \ktcoeyr\ to 823 \ktcoeyr.
This emphasises once more the importance of the fluctuating nature of the annual carbon footprint of
astronomical facilities, driven by the construction of large and impacting facilities.
A snapshot for a given year, as provided in Paper I, does not capture the full dynamics of the GHG
emissions, which only becomes visible once the temporal evolution of facility construction and operations
is taken into account.


\end{document}